\documentclass[accepted]{uai2026} 
                        
\usepackage[american]{babel}

\usepackage{natbib} 
\usepackage{mathtools} 
\usepackage{booktabs} 
\usepackage{tikz} 

\usepackage{amssymb}
\usepackage{braket}
\usepackage{quantikz}
\usepackage{comment}

\usepackage{amsmath}
\usepackage{amsthm}
\usepackage{graphicx}
\usepackage{subcaption}

\newcommand{\X}{\mathcal{X}}

\newcommand{\N}{\mathbb{N}}

\newcommand{\E}{\mathbb{E}}
\newcommand{\ind}{\mathbf{1}}
\newcommand{\norm}[1]{\|#1\|}

\newcommand{\eqdef}{\stackrel{\text{def}}{=}}
\newcommand{\proba}{\mathbb{P}}

\newtheorem{theorem}{Theorem}[section]

\newtheorem{lemma}[theorem]{Lemma}

\newtheorem{definition}[theorem]{Definition}

\newtheorem{remark}[theorem]{Remark}

\title{Benign Overfitting with Quantum Kernels}

\author[1,2]{\href{mailto:<joachim.tomasi@univ-amu.fr>?Subject=Benign overfitting with quantum kernels.UAI 2026 paper}{Joachim Tomasi}}
\author[2]{Sandrine Anthoine}
\author[1]{Hachem Kadri}

\affil[1]{%
    Aix-Marseille Univ\\
    CNRS\\
    LIS\\
    Marseille\\
    France.
}
\affil[2]{%
    Aix Marseille Univ\\
    CNRS\\
    I2M\\
    Marseille\\
    France.
}
  
\begin{document}
\maketitle

\begin{abstract}
Kernel methods compare inputs through feature maps. Quantum kernels follow the same principle: input data are encoded into quantum states, which define \emph{quantum} feature representations in Hilbert spaces.
Kernel values are then obtained by estimating inner products between these states using suitable quantum circuit measurements.
As a result, quantum kernels may be intractable to compute classically\footnotemark while remaining efficiently computable on quantum hardware, potentially leading to a quantum advantage. However, designing effective quantum kernels remains a major challenge. Many quantum kernels, such as the fidelity kernel, suffer from exponential concentration. This results in near-identity kernel matrices that fail to capture meaningful data correlations and lead to overfitting and poor generalization. In this paper, we propose a novel strategy for constructing quantum kernels that achieve good generalization performance, drawing inspiration from benign overfitting in classical machine learning. We introduce the concept of \emph{Local-Global} quantum kernels, which combine two components: a local quantum kernel based on measurements of small subsystems, and a global quantum kernel derived from full-system measurements. To support the effectiveness of the proposed construction, we show theoretically and empirically that Local-Global quantum kernels exhibit benign overfitting.
\end{abstract}

\footnotetext{Here, 'classical' refers to methods or computations that do not rely on quantum computing resources.}
%

\section{Introduction}\label{sec:intro}
    Quantum machine learning (QML) has recently attracted significant attention~\citep{biamonte2017quantum,ciliberto2018quantum,huang2021power,cerezo2022challenges}, particularly as Noisy Intermediate-Scale Quantum (NISQ) devices have become increasingly accessible~\citep{preskill2018quantum}. Variational Quantum Algorithms (VQAs), which operate under a hybrid quantum-classical paradigm, have received considerable interest, mainly due to the possibility of achieving quantum advantage under NISQ constraints~\citep{peruzzo2014variational,farhi2014quantum, du2020expressive}. These algorithms utilize Parameterized Quantum Circuits (PQCs) to compute expectation values of quantum circuit measurements, which are then processed classically to update circuit parameters and optimize cost functions, enabling PQCs to serve as machine learning models~\citep{benedetti2019parameterized,mitarai2018quantum}. 
 
    Quantum kernels, such as quantum fidelity, offer alternative approaches that align with the constraints of NISQ-compatible hardware~\citep{havlivcek2019supervised,schuld2019quantum,schuld2021supervised}. Data are encoded into quantum systems through parameterized unitary gates that maps classical data into quantum states. This mapping, known as a quantum feature map, is the basis of the concept of \textit{quantum kernels}: by encoding classical data into quantum feature states, the quantum fidelity kernel function is defined as the squared magnitude of the inner product between the feature states.
    This can be regarded as a quantum analog of the kernel trick.
    All quantum operations in the quantum kernel space remain linear, akin to classical kernel methods that exploit linear separability in kernel feature spaces~\citep{havlivcek2019supervised,schuld2019quantum}. The kernel trick enables nonlinear decision boundaries in the original data space while avoiding explicit computation of high-dimensional features. 
 
    While VQAs optimize explicit parameterized models, quantum kernels act as their implicit counterparts, leveraging the inner products of embedded quantum states. This duality establishes a theoretical bridge between certain VQA architectures and kernel methods~\citep{schuld2021supervised,huang2021power,jerbi2023quantum}. Quantum kernel methods are founded on the potential to achieve a computational  advantage in feature representations~\citep{mengoni2019kernel}. In this paradigm, certain quantum feature maps are proposed to encode data into high-dimensional Hilbert spaces that are classically intractable but remain efficiently computable on quantum hardware~\citep{liu2021rigorous,kubler2021inductive,havlivcek2019supervised,schuld2019quantum}. For example, as shown in~\citet{liu2021rigorous}, specific quantum circuits can generate feature maps with provable classical intractability under complexity-theoretic assumptions, potentially enabling learning tasks that are beyond the reach of conventional classical methods. Furthermore, recent work has shown that quantum  kernels can approximate broad classes of functions~\citep{gil2024expressivity}. 

    However, despite their computational promise, it remains unclear whether quantum advantages in quantum kernels directly translate into practical improvements in learning performance. One concern is that quantum-enhanced features do not necessarily capture the intrinsic structure of the target learning problems. While these quantum models exhibit high expressivity, they often give rise to challenges such as the exponential concentration of kernel values near zero, resulting in kernel matrices that closely approximate the identity matrix~\citep{thanasilp2024exponential}. Such behavior undermines the model's ability to capture meaningful data correlations and thus limits its generalization capability.

    To achieve generalization in the context of quantum kernels, some research has focused on mitigating overfitting through regularization techniques that avoid mere training set interpolation. For example, methods for tuning kernel bandwidths in quantum kernels and reducing the dimensionality of the quantum feature space have been proposed~\citep{shaydulin2022importance,canatar2023bandwidth,huang2021power,kubler2021inductive}. Another promising direction for mitigating overfitting is to leverage PQCs to learn task-adapted quantum feature maps with the aim of improving generalization performance in quantum kernel methods ~\citep{hubregtsen2022training,rodriguez2025neural}.

    In this work, we study how to enable generalization in interpolating quantum kernel machines, drawing inspiration from recent advances in the  modern classical literature where overparameterized models could generalize well despite achieving near-zero training error.
    Very few studies have examined the generalization abilities of overparameterized quantum models~\citep{gil2024understanding}. \citet{larocca2023theory} have investigated the generalization of overparameterized quantum neural networks (QNNs), while~\citet{kempkes2026double} and~\citet{kamisoyama2026double} have studied the double descent phenomenon in least-squares linear regression within quantum features.  \citet{peters2023generalization} have examined benign overfitting in linear quantum models for uniformly spaced training data.  In our work, we advance this line of research by proposing a framework for quantum kernels that inherently promotes \textit{benign overfitting}. Drawing on the concept of spiky-smooth kernels~\citep{haas2023mind}, we propose a \textit{Local-Global} quantum kernel constructed as a weighted sum of two quantum kernels. One component is derived from a local measurement (i.e., low-dimensional relative to the full quantum embedding space)  that mimics the smooth kernel behavior, while the other comes from a global measurement of the entire quantum feature space, mimicking the spiky component. 
    This Local-Global quantum kernel allows the learned model to have good generalization even when it interpolates training data, giving rise to benign overfitting.

\paragraph{Contributions.} This work introduces benign overfitting as a design principle for quantum kernels. In classical machine learning, benign overfitting is typically studied as a phenomenon explaining how highly overparameterized models can generalize despite interpolating training data~\citep{bartlett2020benign}. In quantum kernel methods, however, interpolation can arise naturally from kernel concentration: as the number of qubits increases, highly expressive quantum kernels may approach the identity matrix~\citep{thanasilp2024exponential}. Rather than treating this concentration solely as a drawback, we show that it can be made useful. Our Local-Global construction combines a spiky global component, which induces localized peaks and enables interpolation, with a smooth local component, which preserves meaningful correlations and supports generalization. This provides, to the best of our knowledge, the first theoretical and empirical characterization of benign overfitting for quantum kernels. From a classical perspective, our analysis extends spiky-smooth benign-overfitting mechanisms beyond shrinking-bandwidth constructions~\citep{haas2023mind} by showing that localization can instead arise from power concentration of the global kernel component. This perspective broadens the scope of benign-overfitting analyses to other families of kernels, including periodic translation-invariant kernels, for which localization is not naturally captured by bandwidth shrinkage.

The main contributions are as follows:
\begin{itemize}
    \item We introduce \emph{Local-Global quantum kernels}, which combine local subsystem and global full-system measurements to obtain a quantum spiky-smooth structure: the global component produces localized peaks, while the local component preserves smooth correlations~(Definition~\ref{de:lgk}).
    \item We prove that, for a class of periodic translation-invariant quantum kernels, ridgeless regression with the Local-Global kernel can interpolate noisy training data while remaining statistically consistent~(Theorems \ref{thm:simplified-quantum-benign2} and \ref{thm:per-quantum-benign}).
    \item We provide quantum circuit implementations of the proposed Local-Global kernel construction, including both fully quantum and hybrid quantum-classical implementations~(Section~\ref{sec:sge_impl}).
    \item We empirically demonstrate on synthetic and real-world datasets that the proposed quantum kernel construction can interpolate the training data while achieving good generalization performance~(Section~\ref{sec:experiments} and Appendix~\ref{app:xp}). We release the code at
\href{https://github.com/jotomasi/benign-overfitting-with-quantum-kernels}{%
\shortstack[l]{%
\texttt{github.com/jotomasi/benign-}\\
\texttt{overfitting-with-quantum-kernels}%
}%
}.
\end{itemize}

\section{Background}\label{sec:background}

We recall here prior results on benign overfitting in kernel regression and quantum kernels that underpin the framework considered in this paper.

\subsection{Benign Overfitting in Kernel Regression}
\label{subsec:backgroundKR}

\label{sec:data-model}
Let \( \mathcal X \subset \mathbb{R}^d \) be a bounded domain and let \( \mu \) be a probability measure on \(\mathcal X \) admitting a density with full support\footnote{ i.e.,  $\forall x \in \mathcal{X}, \; \forall t > 0, \quad \mu(B(x,t)) > 0$, where $B(x,t)$ is the Euclidean ball of radius $t$ centered at $x$. }.
We observe an i.i.d.\ dataset
$\mathcal{D}_n = \{(x_i, y_i)\}_{i=1}^n$,
generated according to
\[
x_i \sim \mu, \qquad y_i = f^\star(x_i) + \varepsilon_i,
\]
where \( f^\star \) is an unknown target function and \( \varepsilon_i \) are independent, zero-mean random variables with finite variance.

Kernel  Regression (KR) extends linear ridge regression by mapping input data $x$ in $\mathcal{X} $ into a high-dimensional Hilbert space $\mathcal{H}_k$ via a feature map $\phi: \mathcal{X} \rightarrow \mathcal{H}_k$. $\mathcal{H}_k$ is a Reproducing Kernel Hilbert Space~(RKHS) defined through a positive definite kernel $k$ satisfying $k(x,z) = \langle \phi(x),\phi(z) \rangle_{\mathcal{H}}$.
Kernel regression seeks a function \( f \in \mathcal{H}_k \) minimizing the regularized empirical risk
\begin{equation}
\mathcal{L}_\lambda(f, \mathcal{D}_n)
:=
\frac{1}{n} \sum_{i=1}^n \bigl(y_i - f(x_i)\bigr)^2
+
\lambda \|f\|_{\mathcal{H}_k}^2,
\label{eq:kernel_ridge}
\end{equation}
where \( \lambda > 0 \) is the ridge regularization parameter.
By the representer theorem, the solution admits the expansion
\[
\hat f(x) = \sum_{i=1}^n \alpha_i k(x_i, x), \quad \text{with } \alpha = (K + \lambda I_n)^{-1} Y,
\]
where \( K \in \mathbb{R}^{n \times n} \) is the kernel matrix (\( K_{ij} = k(x_i, x_j) \))
and $Y=(y_1,\ldots,y_n)^\top$. 
The case \( \lambda = 0 \) is referred to as \emph{ridgeless regression}.
In this case, the kernel regression problem reduces to solving the linear system $K \alpha = Y$.
When \( K \) is not invertible, the ridgeless solution is defined as the minimum-norm interpolant
\[
\alpha = K^{+} Y,
\]
where \( K^+ \) denotes the Moore-Penrose pseudoinverse of \( K \).
If the kernel matrix has sufficiently high rank so that $Y$ lies in its column space, this estimator exactly interpolates the training data.
Throughout the paper, we assume that \( f^\star\) is in \( \mathcal{H}_k \).

\paragraph{From generalization of overparameterized models to benign overfitting in kernel regression}
Overparameterized neural networks can demonstrate strong generalization capabilities, even when they nearly interpolate training data, challenging the classical bias-variance tradeoff. 
This observation has prompted a reevaluation of generalization in modern machine learning~\citep{zhang2021understanding}, leading to various interpretations of the phenomenon.
    One such interpretation is benign overfitting, or harmless interpolation, where models can overfit the training data without compromising their ability to generalize~\citep{bartlett2020benign,muthukumar2020harmless}. In this context, \textit{overfitting} refers to a model trained to (nearly) interpolate the training data, while the terms \textit{benign} or \textit{harmless} describe scenarios in which this overfitting does not negatively impact the model's generalization ability. 
    
    Motivated by the desire to understand this behavior in deep networks, early works explored benign overfitting in simpler models. Notably, ~\citet{bartlett2020benign} studied benign overfitting in overparameterized linear regression, demonstrating that this phenomenon is not limited to deep networks. Furthermore, \citet{mallinar2022benign} refined the classification of overfitting by distinguishing between benign, tempered, and catastrophic regimes, which are based on the behavior of the expected risk of interpolating predictors.
    This phenomenon has also been observed in kernel regression \citep{belkin2019does,liang2020just}, which can be seen as overparameterized linear regression in kernel feature spaces. When the input (or feature) dimension increases with the number of data points, the effective dimension of the kernel matrix grows, and its eigenvalues decay slowly enough that noise gets spread across many less important directions~\citep{bartlett2020benign,tsigler2023benign,bartlett2021deep,hastie2022surprises}. In this context, the minimum-norm interpolant, i.e., the ridgeless regression solution that minimizes its norm, can memorize noise in redundant directions while still capturing the underlying signal. This behavior has been rigorously characterized in linear regression and extended to kernel methods~\citep{tsigler2023benign,zhang2025phase,mallinar2022benign}. In fixed-dimensional settings, some studies showed that benign overfitting can occur under specific conditions, such as constraints on the kernel's eigenvalue spectrum~\citep{pmlr-v235-barzilai24a,pmlr-v235-cheng24g,mallinar2022benign} or on the kernel function form~\citep{haas2023mind}.

    The intuition behind benign overfitting in both linear and kernel regression is provided by the \emph{simple-spiky} decomposition of the minimum-norm interpolant~\citep{bartlett2021deep}. In this view, the learned predictor decomposes into a \emph{simple} component that captures the underlying signal and a \emph{spiky} component helping to interpolate training data without significantly affecting predictions away from the training data~\citep{li2023benign,tsigler2023benign,mallinar2022benign,medvedev2024overfitting,liang2020just}.

    Building on this intuition, \citet{haas2023mind} introduced \emph{spiky-smooth kernels} of the form
    $
    k_{\rho,\gamma}(x,z) = \tilde{k}(x,z) + \rho \, \hat{k}_\gamma(x,z),
    $
    where $\tilde{k}$ is a smooth universal kernel and $\hat{k}_\gamma$ is a spiky kernel, enabling benign overfitting  in fixed-dimensional setting.
    This simple-spiky perspective will guide our construction of quantum kernels.

\begin{table*}[!t]
\centering
\caption{
Classical-quantum correspondence for kernel methods.
$\mathcal X$ is the input space, $\mathcal F$ is a classical feature Hilbert space, and
$\mathcal H_t=(\mathbb C^2)^{\otimes t}$ is the Hilbert space of $t$ qubits. We denote by
$
\mathcal D(\mathcal H_t)
=
\{\rho:\mathcal H_t\to\mathcal H_t \;|\;
\rho=\rho^\dagger,\ \rho\succeq 0,\ \operatorname{Tr}(\rho)=1\}
$
the set of $t$-qubit  states represented as density matrices. In quantum kernel methods, a circuit $U(x)$ encodes a classical input $x$ into a quantum state $\rho_x$. Similarities are then computed using the Hilbert-Schmidt inner product between density matrices.\\[-0.2cm]
}
\label{tab:classical_quantum_correspondence}

\setlength{\tabcolsep}{2.5pt}
\renewcommand{\arraystretch}{1.2}
\begin{tabular}{
@{}
p{0.19\textwidth}
p{0.21\textwidth}
p{0.21\textwidth}
p{0.33\textwidth}
@{}
}
\toprule
\textbf{Concept}
&
\textbf{Classical}
&
\textbf{Quantum}
&
\textbf{Interpretation}
\\
\midrule

Input data
&
\((x,z)\in\mathcal X \times \mathcal X\)
&
\((x,z)\in\mathcal X\times \mathcal X\)
&
Inputs are classical data points in both settings.
\\

Feature space
&
\(\mathcal F\)
&
\(\mathcal D(\mathcal H_t)\)
&
Classical features live in a Hilbert space; quantum features are density matrices acting on the $t$-qubit Hilbert space.
\\

Feature map
&
\(x\mapsto \phi(x)\in\mathcal F\)
&
\(x\mapsto \rho_x\in\mathcal D(\mathcal H_t)\)
&
A feature map sends each input to a feature representation.
\\
Feature representation
&
$\phi(x)$
&
$\rho_x$
&
The quantum state $\rho_x$ plays the role of the feature representation.
\\
Implementation
&
\(\phi:\mathcal X\to\mathcal F\)
&
\(\rho_x=U(x)\rho_{\mathrm{init}}U(x)^\dagger\)
&
We consider quantum feature maps implemented by data-dependent unitary circuits $U(x)$.
\\

Kernel function
&
\(k(x,z)=\langle \phi(x),\phi(z)\rangle_{\mathcal F}\)
&
\(k(x,z)=\operatorname{Tr}[\rho_x\rho_z]\)
&
Both kernels are inner products between feature representations; in the quantum case this is the Hilbert-Schmidt inner product.
\\

\bottomrule
\end{tabular}
\end{table*}

\subsection{Quantum Kernels}\label{subsec:backgroundQK}

\paragraph{Notation}
    
    According to the bra-ket notation adopted in quantum computing, column vectors are denoted as a  ``ket'', $ \ket{\cdot} $, while row vectors are represented as a ``bra'', $ \bra{\cdot} $. These two are dual to each other, with the bra defined as $\bra{\cdot}:= \ket{\cdot}^\dagger,$ where $ \dagger $ denotes the adjoint (or conjugate transpose). The inner product of two states $\psi$ and $\phi$ is written as $ \braket{\psi | \phi} $, and their tensor product is denoted either as $ \ket{\psi\phi} $, $ \ket{\psi}\ket{\phi} $, or equivalently as $ \ket{\psi} \otimes \ket{\phi} $. Also, the $ t $-fold tensor product of a state $ \ket{\psi} $, denoted by $\bigotimes_{i=1}^t\ket\psi$, can be compactly expressed as $ \ket{\psi}^{\otimes t} $ or $ \ket{\psi^t} $.

    Quantum states can also be represented using density matrices. For a pure state $\ket{\psi}$, the density matrix is defined as $\rho = \ket{\psi}\bra{\psi}$. For single-qubit systems the states $\{\ket{0},\ket{1}\}$ define the computational basis of $\mathbb{C}^2$, i.e. $\ket{0}= \begin{pmatrix}1 \quad 0\end{pmatrix}^T$ and  $\ket{1}= \begin{pmatrix}0 \quad1\end{pmatrix}^T$. For $n$-qubit systems, the computational basis is generated by the states $\ket{i}$ for $ i \in \{0,1,\ldots,2^n-1\}$, which corresponds to the tensor product of individual qubit states derived from the bit decomposition of $i$, namely if $i = \sum_{k=0}^{n-1} i_k  2^{k}$ with $i_k \in \{0,1\}$, then $\ket{i} := \ket{i_0} \otimes \dots  \otimes \ket{i_{n-1}}$.
    Quantum observables are represented by Hermitian operators, and the expectation value of an observable $O$ in the state $\rho$ is given by $\braket{O}_\rho = \operatorname{Tr}(\rho O)$. For clarity, the subscript $\rho$ is often omitted when the context is clear. 

    Quantum circuits provide a diagrammatic representation of quantum computations. They consist of a sequence of quantum gates -- unitary operations applied to qubits -- followed by measurements of quantum observables. These circuits offer a structured framework for manipulating quantum states and designing quantum algorithms.
    For a comprehensive introduction to quantum computing and quantum circuits, readers may refer to~\citet{nielsen2010quantum}.
    
    \begin{figure}[t]
        \centering
        \begin{quantikz}
    \lstick{\(\rho_{\mathrm{init}}\)}&\qwbundle{t} & \gate{U(x)} & \gate{U^\dagger(z)} & \meter{\rho_{\mathrm{init}}} & \setwiretype{n} \mathrm{Tr}[\rho_x\rho_z]
    \end{quantikz}
    \caption{
    Quantum circuit for evaluating the quantum kernel. 
    The $t$-qubit register is initialized in the reference state $\rho_{\mathrm{init}}$. 
    Given inputs $x$ and $z$, the circuit prepares 
    $\sigma_{x,z}=U^\dagger(z)U(x)\rho_{\mathrm{init}}U^\dagger(x)U(z)$. 
    Measuring $\rho_{\mathrm{init}}$ yields the expectation value $\mathrm{Tr}[\sigma_{x,z}\rho_{\mathrm{init}}]
    = \mathrm{Tr}[\rho_x \rho_z],$
    by cyclicity of the trace, where 
    $\rho_x = U(x)\rho_{\mathrm{init}}U^\dagger(x)$. 
    This equals the kernel value $k(x,z)$. 
    For pure-state encoding, this reduces to 
    $k(x,z)=|\langle \phi_x | \phi_z \rangle|^2$.
    }
    \label{fig:circqk}
    \end{figure}

    \paragraph{Quantum encoding}  A quantum machine learning algorithm needs data in the form of quantum states. So classical data should be first encoded into quantum states, i.e., the transformation of a classical data $x$ to a quantum state $\ket{\phi_x}$. Most of the interest in quantum kernels comes from the observation that encoding classical data into a quantum computer defines an explicit feature representation of the data.
    Consider a quantum feature map that encodes a data point $x$ into a quantum state represented by the density matrix $\rho_x$. A quantum kernel $k(x,z)$ is defined as the  Hilbert-Schmidt inner product between the states $\rho_x$ and $\rho_z$, given by
    \begin{equation}
    \label{eq:qk}
        k(x,z) = \mathrm{Tr}[\rho_x \rho_z].
    \end{equation}
Figure~\ref{fig:circqk} depicts a circuit implementing the quantum kernel where each data point $x$ is encoded as a density matrix of the form $\rho_x = U(x)\,\rho_{\mathrm{init}}\,U^\dagger(x)$.

In the case where $\rho_{\mathrm{init}}$ is a pure state, for example $\rho_{\mathrm{init}} =\ket{0^t}\bra{0^t} $ the quantum kernel in \eqref{eq:qk} simplifies to
    \begin{equation}
        k(x,z) = |\braket{\phi_x| \phi_z}|^2 = |\braket{0^t | U^\dagger(z) U(x) | 0^t}|^2. 
        \label{eq:qfk}
    \end{equation}
    This corresponds to measuring the fidelity between the quantum states $\ket{\phi_x}$ and $\ket{\phi_z}$, making this kernel known as the quantum fidelity kernel~\citep{schuld2019quantum,mengoni2019kernel}. 
Table~\ref{tab:classical_quantum_correspondence} summarizes the classical-quantum correspondence for kernel methods.

Recent studies analyzed generalization error bounds for learning with quantum fidelity kernels and the results appear to be negative~\citep{huang2021power,kubler2021inductive}. The expressive power of quantum models can hinder generalization. Finding suitable quantum kernels is therefore challenging, since highly expressive quantum embeddings may induce a geometry that is poorly aligned with the learning task~\citep{huang2021power}; in particular, kernel values may concentrate as the number of qubits increases~\citep{thanasilp2024exponential}. In other words, when using a large number of qubits, the kernel matrix 
gets close to the identity matrix, resulting in overfitting and poor generalization performance~\citep{suzuki2024quantum}. To tackle this issue, \citet{canatar2023bandwidth} proposed the use of a bandwidth parameter for quantum fidelity kernels, while \citet{huang2021power,kubler2021inductive} considered constructions based on reduced density matrices.

\section{Benign Overfitting with Local-Global Quantum Kernels}\label{sec:boqk}

In this section, we introduce our main contribution: Local-Global quantum kernels, which are quantum analogs of spiky-smooth kernels that naturally promote benign overfitting. These kernels explicitly decompose into smooth and spiky components while remaining compatible with the quantum circuit-based measurement formalism for kernel evaluation.

\subsection{Local-Global Quantum Kernel}\label{sec:lgqk}
A central challenge in quantum kernel methods is the trade-off between expressivity and generalization as the number of qubits increases. When kernels are constructed from similarities between full quantum feature states, inner products tend to concentrate around zero in high-dimensional Hilbert spaces and consequently yield kernel matrices that approach the identity~\citep{thanasilp2024exponential}. Such \emph{spiky} kernels enable interpolation of training data but may harm generalization. In contrast, kernels constructed from reduced quantum states of a subsystem retain a smoother structure and thereby preserve nontrivial correlations between distinct inputs and can better capture the underlying signal~\citep{huang2021power}.

Motivated by this phenomenon and by the spiky-smooth decomposition underlying benign overfitting in classical kernel regression~\citep{haas2023mind}, we introduce a \emph{Local-Global quantum kernel} that combines:
(i) a \emph{global} kernel, defined from similarities between full quantum feature states and exhibiting concentration effects, and
(ii) a \emph{local} kernel, defined from similarities between reduced quantum feature states on a subsystem and inducing smoother behavior.
This construction combines a spiky and a smooth kernel component and should support both expressivity and generalization.

\begin{definition}[Local–Global Quantum Kernel]
    \label{de:lgk}
    Let $U(x)$ be a unitary operator acting on $t$ qubits. Define the local and global quantum feature maps
    \[
    \rho_x^{L} := U(x)\Bigl(L_s \otimes \frac{1}{2^{t-s}} I_{t-s}\Bigr)U^\dagger(x), 
    \rho_x^{G} := U(x) G_t U^\dagger(x),
    \]
    where $L_s$ is a fixed pure quantum state (rank-one projector) on $s < t$ qubits and $G_t$ is a fixed pure quantum state on $t$ qubits. The \emph{local–global quantum kernel} is defined as
    \begin{equation}\label{eq:lgqk}
        k_{LG}(x,z) = \lambda_L\, k_{L}(x,z) + \lambda_G\, k_{G}(x,z),
    \end{equation}
    with
    \[
        k_{L}(x,z) := \mathrm{Tr}\!\big[\rho_x^{L}\rho_z^{L}\big], 
        \qquad 
        k_{G}(x,z) := \mathrm{Tr}\!\big[\rho_x^{G}\rho_z^{G}\big],
    \]
    where $\lambda_L, \lambda_G \ge 0$ are scalar weights.
\end{definition}

The terminology \emph{local} and \emph{global} refers to the structure of the quantum measurement underlying the kernel evaluation. 
The global kernel compares full $t$-qubit quantum feature states, whereas the local kernel compares reduced states obtained by tracing out $t-s$ qubits and retaining only a subsystem of size $s$.

More formally, the \emph{global kernel} is constructed from $G_t$ which is a rank-one projector, i.e., $G_t = \ket{\psi}\bra{\psi}$. For instance, choosing
\[
G_t = \rho_{\mathrm{init}}^{G} = \ket{0^t}\bra{0^t}
\]
yields the standard quantum fidelity kernel as defined in~\eqref{eq:qfk}.
In contrast, the \emph{local kernel} is obtained by initializing only a subset of $s<t$ qubits in a pure state, while the remaining qubits are maximally mixed, i.e.,
\begin{equation}
\rho_{\mathrm{init}}^{L}
=
L_s \;\otimes\; \frac{1}{2^{t-s}} I_{t-s}.
\label{eq:local_init}
\end{equation}
Letting $\rho_x^L = U(x)\rho_{\mathrm{init}}^{L}U^\dagger(x)$ and using the cyclic property of the trace together with properties of the partial trace~\citep{filipiak2018properties}, we obtain
\begin{align}
\label{eq:sigma_xz}
k_L(x,z)
&= \mathrm{Tr}\!\left[ \rho_x^L\rho_z^L \right] \nonumber \\
&= \mathrm{Tr}\!\left[ U^\dagger(z)U(x)\rho_{\mathrm{init}}^{L}U^\dagger(x)U(z)\rho_{\mathrm{init}}^{L} \right] \nonumber \\
&= \mathrm{Tr}\!\left[ \sigma_{x,z}\,\rho_{\mathrm{init}}^{L} \right]
= \frac{1}{2^{t-s}}\mathrm{Tr}\!\left[ \widetilde{\sigma}_{x,z} L_s \right],
\end{align}
where $\sigma_{x,z} := U^\dagger(z)U(x)\rho_{\mathrm{init}}^{L}U^\dagger(x)U(z)$ and $\widetilde{\sigma}_{x,z}$ denotes its partial trace over the last $t-s$ qubits.
This reformulation makes explicit that $k_L$ depends only
on local measurements performed on the first $s$ qubits, while the remaining qubits are traced out and do not affect the kernel.
As a result, $k_L$ typically yields smoother kernel matrices with non-negligible off-diagonal entries.

The key intuition behind our construction is that
as the number of qubits $t$ increases, the global kernel $k_G$ becomes increasingly spiky, whereas the local kernel $k_L$ remains smooth. The local–global kernel combines these two regimes: the local component supports generalization when the model class becomes highly expressive with increasing qubits, whereas the global component introduces narrow spikes that promote the interpolation of training data.
This realizes a quantum analogue of the spiky–smooth decomposition of~\citet{haas2023mind} and provides a mechanism for benign overfitting with quantum kernels.

\subsection{Separable Global Encoding}
\label{sec:sge}

     \begin{figure}[t]
         \centering
\includegraphics[scale=0.31]{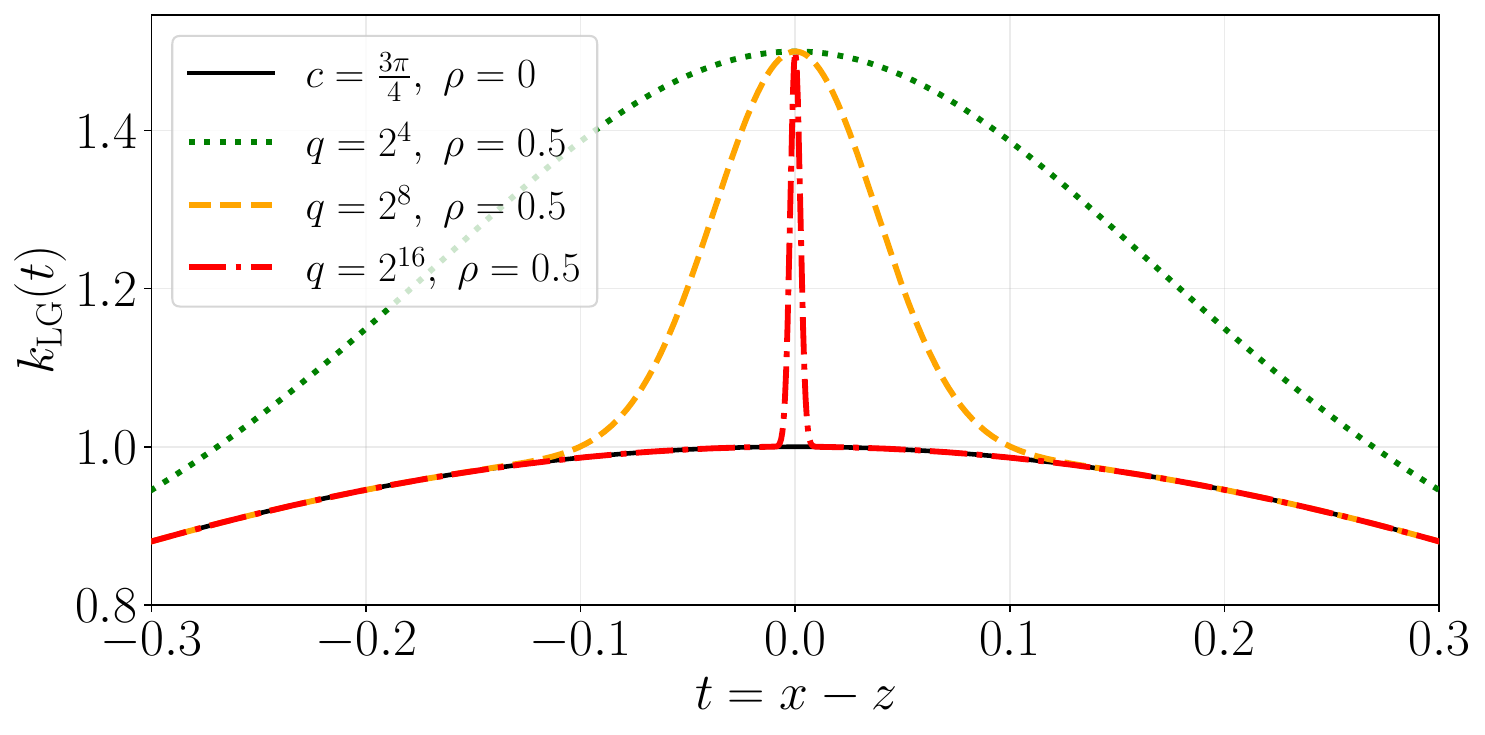}
         \caption{Local-global kernels with $k_{LG}(x,z) =\cos^2(\frac{c(x-z)}{2}) + \rho\cos^{2q}(\frac{c(x-z)}{2})$, where $c=\frac{3 \pi}{4}$, $\rho = 0.5$ and $q = 2^4,2^8,2^{16}$. The kernel $k(x,z) = \cos^2(\frac{c(x-z)}{2})$ is the fidelity quantum kernel obtained by angle quantum encoding~\citep{canatar2023bandwidth}.}
         \label{fig:spsm}
     \end{figure}

To analyze the local–global quantum kernel in a simple and analytically tractable setting, we introduce a structured quantum encoding scheme, which makes the spiking mechanism explicit and enables a precise study of benign overfitting. In particular, this construction clarifies how the choice of the subsystem size $s$, the total number of qubits $t$, and the encoding unitary $U(\cdot)$ controls the smoothness of the local kernel and the spikiness of the global kernel.

Let $s \ge 1$ and consider that 
$L_s= \ket{0^s}\bra{0^s}$. Suppose that the total number of qubits satisfies $t = q\,s$ for some integer $q \ge 1$. We assume that both the encoding unitary and the initial state of the global kernel factorize across $q$ blocks of $s$ qubits, i.e., 
\[
U(x) = V_s(x)^{\otimes q}, 
\qquad 
\rho_{\mathrm{init}}^G := \ket{0^t}\bra{0^t} = (\ket{0^s}\bra{0^s})^{\otimes q},
\]
where $V_s(x)$ acts on $s$ qubits. 
We refer to this scheme as \emph{Separable Global Encoding}.

Under the separable global encoding scheme, the reduced density matrix of the local quantum feature map becomes
\[
\tilde{\rho}_x^L := \mathrm{Tr}_{t-s:t}[\rho^L_x] = V_s(x)\ket{0^s}\bra{0^s}V_s^\dagger(x),
\]
which defines a \emph{base kernel} on $s$ qubits
\begin{equation}
   k_B(x,z) := \mathrm{Tr}[\tilde{\rho}_x^L \tilde{\rho}_z^L] = |\braket{0^s|V_s(x)V_s^\dagger(z)|0^s}|^2.
   \label{eq:bk}
\end{equation}
This base kernel is itself a quantum fidelity kernel, acting on a smaller subsystem than the global kernel and captures low-dimensional structure of the data.
 
Using the tensor structure and linearity of the trace, the local kernel reduces to
\[
k_L(x,z) := \mathrm{Tr}[\rho_x^L\rho_z^L] = \frac{1}{2^{t-s}}\, k_B(x,z).
\]
Thus, the local kernel is proportional to the base kernel and depends only on the reduced $s$-qubit subsystem.

Similarly, the global kernel factorizes over the $q$ blocks, i.e.,
\[
k_G(x,z) := \mathrm{Tr}[\rho_x^G \rho_z^G] = \mathrm{Tr}[(\tilde{\rho}_x^L \tilde{\rho}_z^L)^{\otimes q}] = k_B(x,z)^q.
\]
The global kernel amplifies the spikiness through the exponent $q$, which grows with the total number of qubits relative to the subsystem size.
The local–global kernel in Definition~\ref{de:lgk} then reduces to
\begin{equation}
    \label{eq:lgqk_sepenc}
    k_{LG}(x,z) = \tilde\lambda_L k_B(x,z) + \lambda_G k_B(x,z)^q, 
\end{equation}
where $\tilde\lambda_L = \frac{\lambda_L}{2^{t-s}}$.

The Local-Global quantum kernel in this case includes a parameter $q$, which determines the kernel's degree and serves as a tuning parameter for controlling the behavior of the global component.
     Figure~\ref{fig:spsm} illustrates this idea.  The Local-Global kernel closely follows the local component but deviates in specific narrow regions. As the parameter $q$ increases, these deviations become more localized, reinforcing the analogy with spiky-smooth kernels.

\subsection{Quantum Circuit implementation}
\label{sec:sge_impl}

We now describe how the local–global quantum kernel defined in~\eqref{eq:lgqk_sepenc} can be implemented on quantum hardware.

\textit{Fully quantum implementation.}  
The reduced operator $\tilde{\sigma}_{x,z}$ associated with the local kernel, see~\eqref{eq:sigma_xz}, satisfies
\begin{equation}
\tilde{\sigma}_{x,z} := \mathrm{Tr}_{s+1:t}[\sigma_{x,z}] 
= V_s^\dagger(z) V_s(x) \, \ket{0^s}\bra{0^s} \, V_s^\dagger(x) V_s(z).
\end{equation}
Using the factorization properties of the trace over tensor products
and the fact that $ \mathrm{Tr}[\tilde\sigma_{x,z}]  = 1$, the local–global kernel can be expressed as
\begin{align}
k_{LG}(x,z) 
&= \tilde{\lambda}_L \mathrm{Tr}[\tilde{\sigma}_{x,z}\ket{0^s}\bra{0^s}] 
 + \lambda_G \mathrm{Tr}[\tilde{\sigma}_{x,z}\ket{0^s}\bra{0^s}]^q \notag \\
&= \mathrm{Tr}[\tilde{\sigma}_{x,z}^{\otimes q} \, O_{LG}],
\end{align}
where the measurement operator $O$ is defined as
    $O = \tilde \lambda_L O_L + \lambda_G O_G$ with $O_L = \ket{0^s}\bra{0^s} \otimes I_s^{\otimes q-1}$ and $ O_G= (\ket{0^s}\bra{0^s})^{\otimes q}$.

This decomposition shows that $k_{LG}$ can be computed entirely through quantum measurements on $t$ qubits using a weighted sum of local and global observables. Figure~\ref{fig:circuit1} illustrates the corresponding quantum circuit.

\textit{Hybrid classical-quantum implementation.}  
A more resource-efficient approach exploits the separable structure of the kernel. Since $k_{LG}(x,z)$ is fully determined by the base kernel $k_B(x,z)$, we can compute $k_B(x,z)$ using a quantum circuit on only $s$ qubits, and then classically compute the kernel via~\eqref{eq:lgqk_sepenc}.
This hybrid scheme avoids preparing $q$ tensor copies and reduces the required quantum resources from $t=qs$ qubits to just $s$ qubits.  Figure~\ref{fig:circuit_cq} shows this  implementation.

\begin{figure}[t]
        \centering
        \begin{quantikz}
            \lstick{$\ket{0}$} & \qwbundle{t} & \gate{V_s(x)^{\otimes q}} & \gate{V_s^\dagger(z)^{\otimes q}} & \meter{O} & \setwiretype{n}
        \end{quantikz}
        \caption{A quantum circuit for computing the Local-Global kernel $k_{LG}= \braket{O}$ with
    $O = \tilde \lambda_L O_L + \lambda_G O_G$, $O_L = \ket{0^s}\bra{0^s} \otimes I_s^{\otimes q-1}$ and $ O_G= (\ket{0^s}\bra{0^s})^{\otimes q}$.
        }
        \label{fig:circuit1}
    \end{figure}

    \begin{figure}[t]
        \centering
        \vspace{-0.13cm}
        \begin{quantikz}
            \lstick{$\ket{0}$} & \qwbundle{s} & \gate{V_s(x)}               &  \gate{V^\dagger_s(z)} & \meter{\ket{0^s}\bra{0^s}} & \setwiretype{n} 
        \end{quantikz}
        \caption{A quantum circuit for computing the base (local) kernel $k_B=\braket{\ket{0^s}\bra{0^s}}$. The Local-Global quantum kernel $k_{LG}(x,z)$ is then obtained from $k_B$ 
        using~\eqref{eq:lgqk_sepenc}.}
        \label{fig:circuit_cq}
    \end{figure}

\begin{figure*}[t]
    \centering
\includegraphics[width=0.95\linewidth]{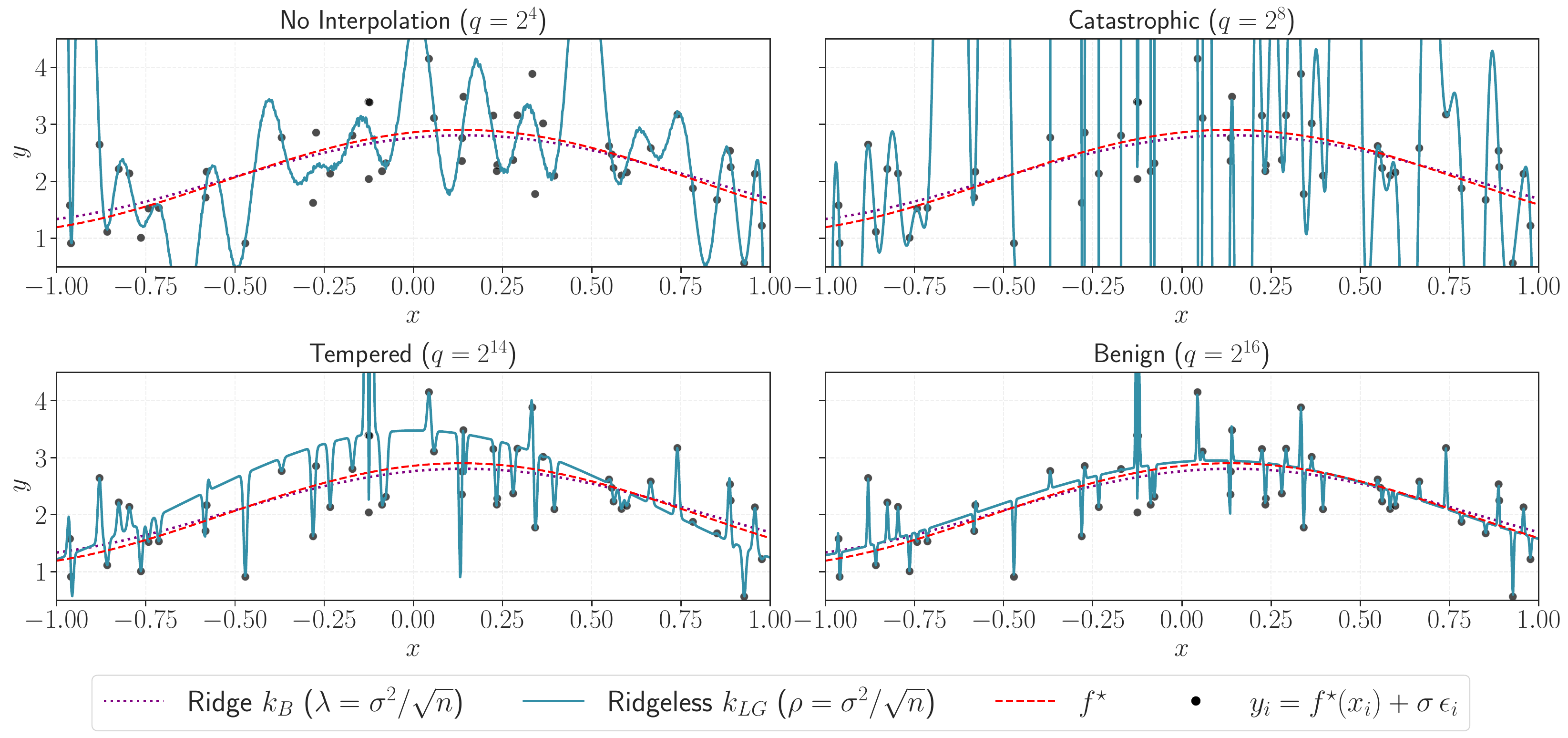}
    \caption{Ridgeless regression using the Local-Global kernel $k_{LG}(x,z)=k_B(x,z)+\rho k_B(x,z)^q$ with different values of $q$.}
    \label{fig:spsmreg}
\end{figure*}

\subsection{Theoretical Analysis}\label{subsec:theory}

We consider \emph{Local-Global quantum kernels} arising from the global separable encoding introduced in Section~\ref{sec:sge}, for which the kernel admits the decomposition
given in~\eqref{eq:lgqk_sepenc}. Throughout this section, we fix $\tilde{\lambda}_L = 1$ and $\lambda_G = \rho$.

The objective here is to demonstrate that \emph{quantum kernels can be constructed to yield interpolating estimators that nevertheless remain statistically consistent}, thereby exhibiting \emph{benign overfitting} in the sense of~\citet{mallinar2022benign,haas2023mind}. Specifically, we show that for a broad class of quantum fidelity kernels, the associated Local-Global kernel induces a ridgeless estimator that interpolates the training data while remaining consistent.
For clarity, we present the result for the cosine quantum kernel as base kernel
\begin{equation}
\label{eq:cos}
   k_B(x,z) = \prod_{i=1}^d \cos^2\!\left(\tfrac{c(x^{(i)} - z^{(i)})}{2}\right), 
\end{equation}

which arises from angle encoding~\citep{canatar2023bandwidth} (for more details, see Appendix~\ref{subsec:angleencoding}). This concrete instance shows the essential mechanism underlying benign overfitting in the Local-Global quantum kernel construction.

\begin{theorem}[Benign overfitting for the cosine quantum kernel]
\label{thm:simplified-quantum-benign2}
Let $k(x,z)$
be the  quantum fidelity kernel  defined in~\eqref{eq:cos}. 
Assume that the data are generated according to the model of Section~\ref{sec:data-model}, and that the input distribution $\mu$ admits a density $g \in L^1(\mathbb{R}^d) \cap L^p(\mathbb{R}^d)$ for some $p > 1$. 
Let $\alpha > 0$ and $0 < \alpha_0 < 1$.

Let $\hat f_{\rho,q,n}$ denote the ridgeless regression estimator trained on $n$ samples with the Local-Global kernel $k + \rho k^q$. 
Suppose that the real positive sequences $(q_n)_n$ and $(\rho_n)_n$ satisfy
\[
n^{\alpha_0} \rho_n \xrightarrow[n \to \infty]{} 0, 
\qquad 
n \rho_n \xrightarrow[n \to \infty]{} \infty,
\]
\[
q_n \ge C n^\xi \ln n,
\qquad 
C > 6+\alpha,
\qquad 
\xi > \tfrac{2(2+\alpha)p}{d(p-1)}, 
\]
and  with high probability the estimator interpolates the data:
$
\hat f_{\rho_n,q_n,n}(x_i) = y_i, \quad \forall i.
$

Then the sequence $\{\hat f_{\rho_n,q_n,n}\}_{n}$ is consistent in probability: 
\[
\mathbb{P}\!\left(
\E_x\!\left[\big(\hat f_{\rho_n,q_n,n}(x) - f^*(x)\big)^2\right] 
> \varepsilon
\right)
\xrightarrow[n \to \infty]{} 0,
\quad \forall \varepsilon > 0.
\]
\end{theorem}

Theorem~\ref{thm:simplified-quantum-benign2} shows that the cosine quantum fidelity kernel admits a Local-Global construction that interpolates the training data while remaining statistically consistent. The sequence $(\rho_n)$ ensures the consistency of the underlying kernel ridge estimator based on $k$, while the sequence $(q_n)$ concentrates the global component $k^{q_n}$ increasingly tightly around the training inputs. This enables exact interpolation of the sample without degrading generalization performance. Consequently, consistency is achieved in the ridgeless regime.

This phenomenon is not specific to the cosine quantum kernel. In Appendix~\ref{subsec:hamiltonian}, we show how to build periodic, translation-invariant quantum fidelity kernels. We then show in Appendices~\ref{app:setting} to~\ref{app:proof} that this whole class of quantum kernels exhibits the same properties as the cosine kernel.

\paragraph{Sketch of proof.}
Benign overfitting requires two properties: interpolation of the training data and consistency of the estimator.

Under the data model of Subsection~\ref{sec:data-model}, the regression function satisfies $f^* \in \mathcal H_k$. Since
$
\mathcal H_k \subseteq \mathcal H_{k + \rho_n k^{q_n}},
$
the learning problem is well specified for both kernels. Consistency therefore reduces to showing
\[
\E_x\!\left[\big(\hat f_{\rho_n,q_n,n}(x) - f^*(x)\big)^2\right] \underset{n\to\infty}\longrightarrow 0.
\]
Let $\hat f_{\rho_n,n}$ denote the kernel ridge estimator with kernel $k$ and regularization parameter $\rho_n$. We have
 \begin{align}
\E_x[(\hat f_{\rho_n,q_n,n}(x)&-f^*(x))^2]  \le 2
\E_x[(\hat f_{\rho_n,n}(x)-f^*(x))^2] \nonumber\\
 +2 &\E_x[(\hat f_{\rho_n,q_n,n}(x)-\hat f_{\rho_n,n}(x))^2]. \label{eq:consistency-decomposition}
\end{align} 

The first term on the right-hand side of \eqref{eq:consistency-decomposition} corresponds to classical kernel ridge regression with kernel $k$ and vanishing regularization. Standard results guarantee its consistency under the stated conditions on $\rho_n$~\citep[chapter 7]{bach2024learning}.

The second term on the right-hand side of \eqref{eq:consistency-decomposition} quantifies the deviation between the ridgeless estimator based on $k + \rho_n k^{q_n}$ and the ridge estimator based on $k$. Following the proof technique of~\citet{haas2023mind}, the key observation is that $k^q$ acts as a \emph{highly localized} kernel: for any fixed $(x,z)$ with $0 \le k(x,z) < 1$, we have $k(x,z)^q \to 0$ as $q \to \infty$. As a result, for large $q_n$, the Hadamard power (element-wise product) $K^{\odot q_n}$ becomes diagonally dominant, and consequently the matrix $K + \rho_n K^{\odot q_n}$ is close to $K + \rho_n I_n$.

To formalize this, we introduce the $\delta$-neighborhood $S(\delta)$ of the spike set
$Z = \{(x,z) \in \mathcal X^2 : k(x,z) = 1\}$ (see Appendix~\ref{app:auxres}). Choosing $\delta_n \to 0$ sufficiently slowly ensures that with high probability no pair of distinct training points lies in $S(\delta_n)$ (Lemma~\ref{lem:min-dist}). Outside $S(\delta_n)$, the kernel $k$ is uniformly bounded away from $1$ by
$m(\delta_n) := \sup_{(x,z) \in S(\delta_n)^c} k(x,z) < 1$, 
so that for $i \neq j$,
\[
k(x_i, x_j)^{q_n} \le m(\delta_n)^{q_n}.
\]
Choosing \(q_n\) sufficiently large relative to \(m(\delta_n)\) ensures that
the second term on the right-hand side of
\eqref{eq:consistency-decomposition} vanishes.
Since both terms vanish, consistency in probability follows. 
A complete proof of the theorem for periodic translation-invariant quantum kernels is provided in Appendix~\ref{app:proof}, while Appendix~\ref{app:cos} details its specialization to the cosine quantum kernel.

\paragraph{Comparison with~\citet{haas2023mind}.} \label{par:comparison}
In the spiky--smooth framework of~\citet{haas2023mind}, the smooth component is chosen to be universal, while the spiky component admits a bandwidth parameter that controls localization around the diagonal. This shrinking-bandwidth mechanism simplifies the analysis of interpolation and consistency.
Such a construction is not available for quantum fidelity kernels. For instance, kernels of the form
$
k_\gamma(x,z) = \cos^2\!\left(\frac{x-z}{\gamma}\right)
$ 
are periodic and do not converge as $\gamma \to 0$. The Local-Global construction introduced here circumvents this limitation: the global component $k^q$ induces localization through \emph{power concentration} rather than bandwidth shrinkage, enabling benign overfitting without altering the  quantum encoding.

\section{Numerical Experiments}
\label{sec:experiments}

\begin{figure}[t]
        \centering
        \includegraphics[width=1\linewidth]{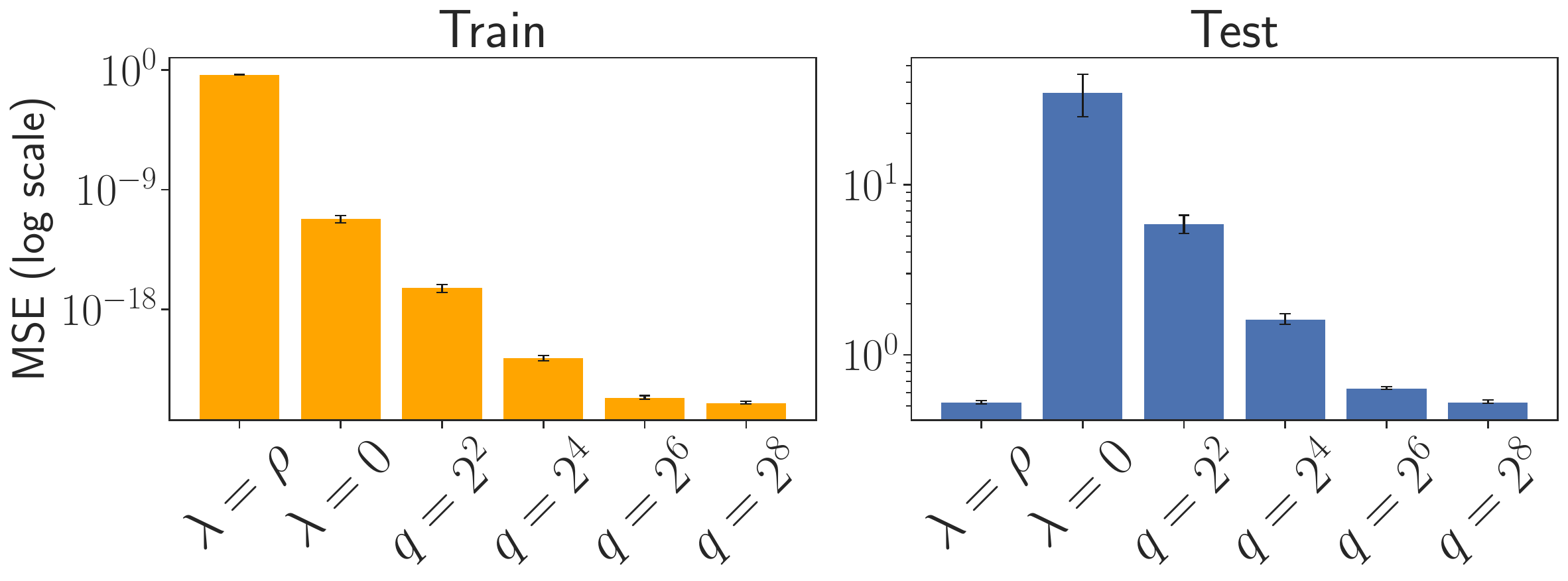}
\caption{
Regression with synthetic data ($n_{\text{train}}=500$, $d=7$). Train and test MSE for $k_B$ (ridgeless $\lambda=0$ \& ridge $\lambda=\rho$) and  $k_{LG}$~(ridgeless) with varying $q$.
}
        \label{fig:synthetic_d7in}
\end{figure}

\begin{figure}
    \centering
    \includegraphics[width=1\linewidth]{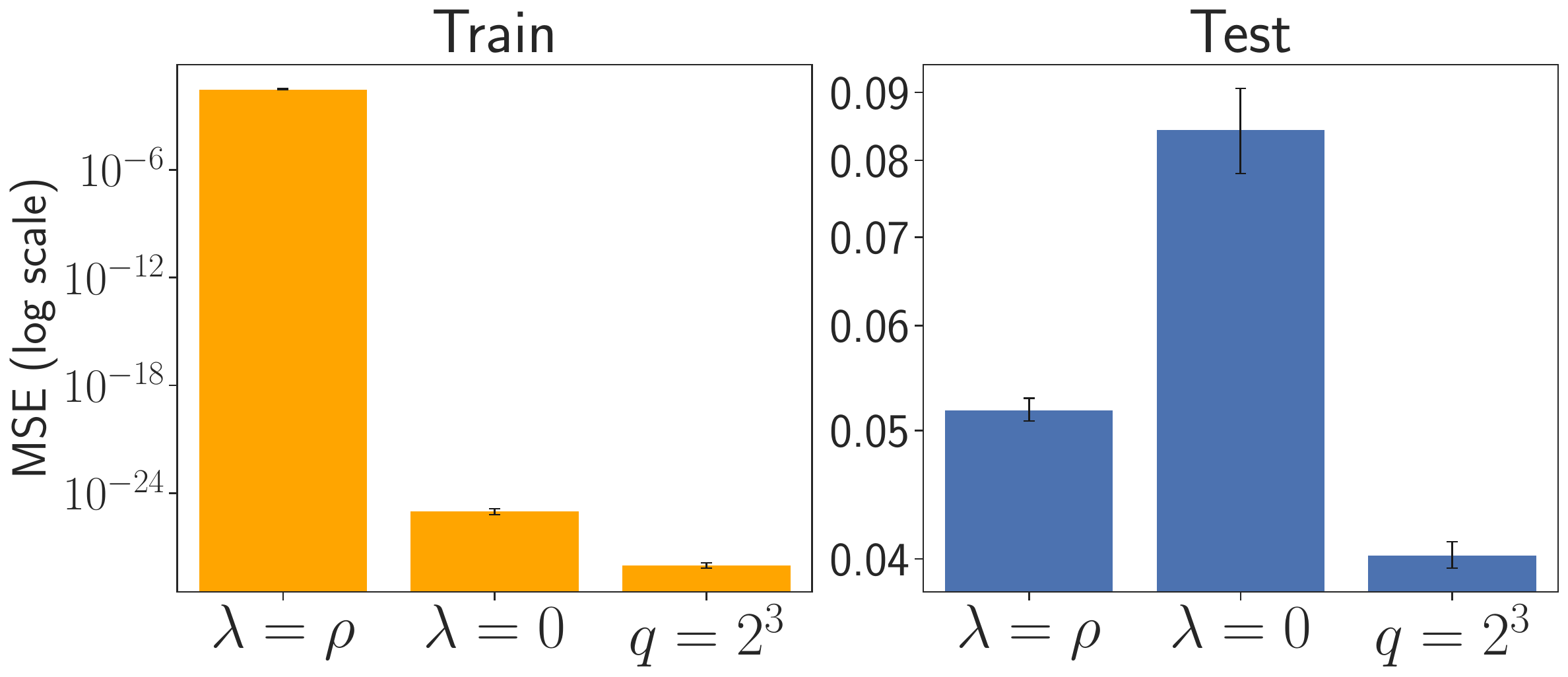}
\caption{Regression with Airfoil dataset ($n_{\text{train}}=500$, $d=20$). Train and test MSE for $k_B$ (ridgeless $\lambda=0$ \& ridge $\lambda=\rho$) and  $k_{LG}$~(ridgeless).
}
    \label{fig:airfoiln500bestq}
\end{figure}

\begin{figure}[t]
        \centering
        \includegraphics[width=1\linewidth]{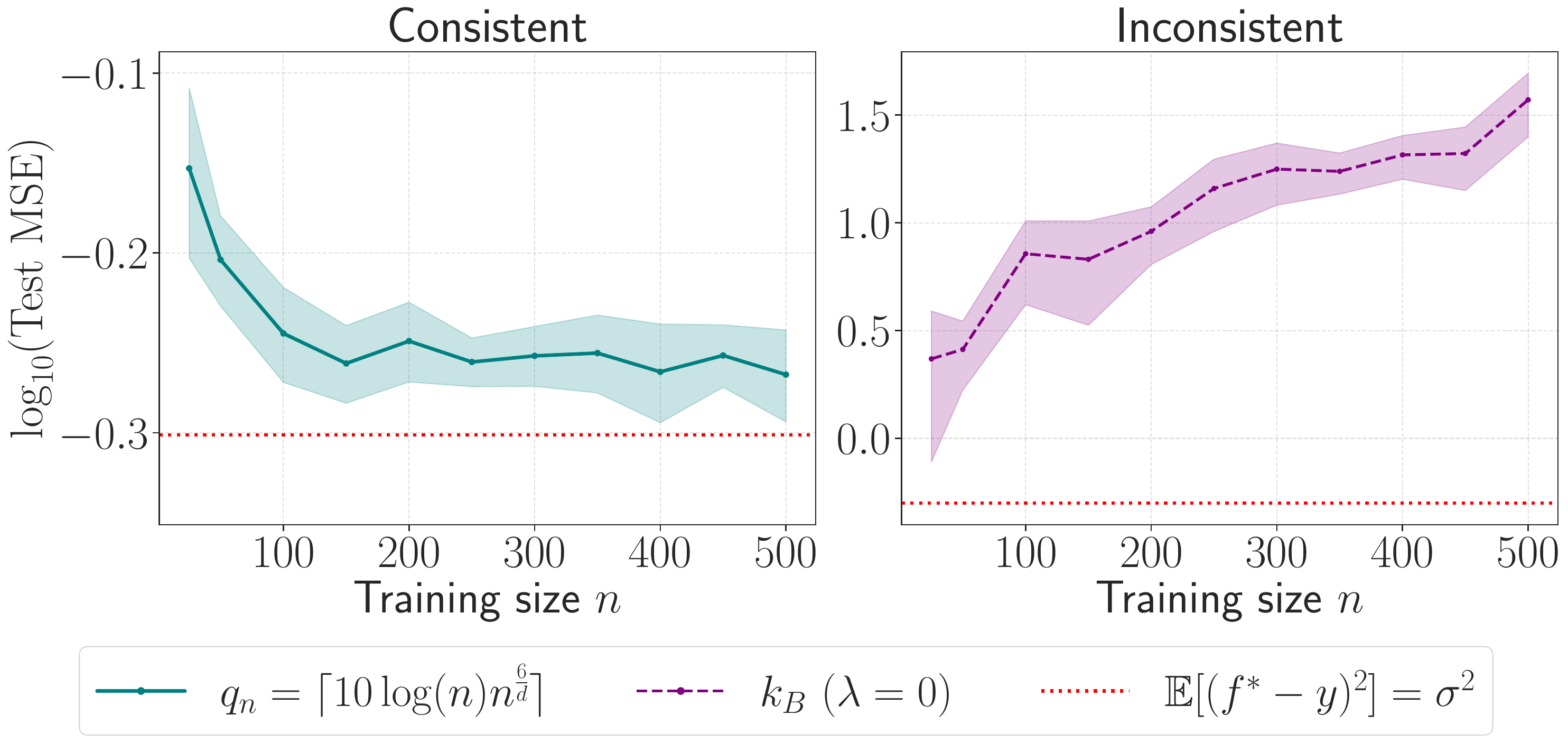}
\caption{Consistency on synthetic data for ridgeless $k_{LG}$ with $\rho_n = \sigma^2/\sqrt{n}$ and $q_n = \lceil 10 \log(n) n^{6/d} \rceil$. The test error converges to the noise level $\sigma^2$. Ridgeless regression with $k_B$ ($\lambda=0$) is inconsistent.
}
        \label{fig:consistency}
\end{figure}

We evaluate the proposed Local-Global kernel construction~\eqref{eq:lgqk_sepenc} with $\tilde{\lambda}_L = 1$ and $\lambda_G = \rho$  with the cosine quantum kernel $k_B$ defined in~\eqref{eq:cos}. 
We compare ridgeless regression with the Local-Global kernel $k_{LG}$ to regression with the kernel $k_B$, both in the ridgeless regime ($\lambda=0$) and with ridge regularization ($\lambda=\rho$).
Experiments are conducted on synthetic and real-world datasets, with performance measured by mean squared error (MSE) on independent test sets and averaged over multiple random splits.
Further details on the datasets and experimental setup are provided in Appendix~\ref{app:xp}.

\paragraph{Illustration of overfitting regimes in 1D} 
We first illustrate the behavior of the Local-Global kernel ridgeless regression in one dimension as the exponent $q$ varies. The target function is constructed as a linear combination of the base kernel $k_B$ evaluated on uniformly sampled points in 
$[-1,1]$. 
Figure~\ref{fig:spsmreg} reveals four distinct regimes as $q$ increases: 
(top left) for small $q$, the estimator fails to interpolate and exhibits large oscillations (poor bias-variance tradeoff); 
(top right) for intermediate $q$, interpolation is achieved but strong oscillations between samples lead to severe overfitting; 
(bottom left) for larger $q$, interpolation becomes spatially localized while predictions remain smooth away from the data; 
(bottom right) for very large $q$, the interpolating predictor remains close to $f^\star$, illustrating benign overfitting. 
This 1D experiment exposes the full overfitting phase diagram induced by quantum kernel powering in the Local-Global construction.

\paragraph{Evaluation on synthetic and real-world data}
We next evaluate the Local-Global kernel on a synthetic regression task (see Figure~\ref{fig:synthetic_d7in}) and on the Airfoil real-world dataset~(see Figure~\ref{fig:airfoiln500bestq}); additional results on other datasets are provided in Appendix~\ref{app:xp}. Ridgeless 
regression with the quantum kernel $k_B$ interpolates the training data yet fails to generalize. The Local-Global kernel preserves interpolation while maintaining strong generalization, and in some cases even outperforms ridge regression.
This improvement does not imply that interpolation is always preferable to regularization. The two estimators search different hypothesis spaces: ridge regression with the local kernel in $\mathcal H_L$, and ridgeless regression with the Local-Global kernel in the larger space $\mathcal H_{LG}$, with $\mathcal H_L\subseteq\mathcal H_{LG}$. When the target is better approximated in $\mathcal H_{LG}$, the Local-Global estimator can improve test performance while still interpolating the training data.

\paragraph{Consistency}
Finally, we illustrate the consistency of the Local-Global estimator on the synthetic data from the previous experiment as the training size $n$ increases. Figure~\ref{fig:consistency} shows that the test MSE of the Local-Global kernel decreases and converges to the noise level $\sigma^2$, which is in agreement with Theorem~\ref{thm:simplified-quantum-benign2}. Ridgeless regression with the  cosine quantum kernel clearly exhibits inconsistency.

\section{Conclusion}
We introduced a novel approach to quantum kernel construction, called \textit{Local-Global} kernels,
which combines local and global measurements to enable benign overfitting in quantum machine learning. By leveraging separable global encoding, the kernel can be naturally implemented on a quantum device while achieving strong generalization.
Our theoretical results establish conditions under which ridgeless regression with this kernel achieves benign overfitting and consistency.
This shows that carefully designed quantum features can interpolate data without compromising predictive performance. Empirical evaluations on both synthetic and real-world datasets confirm these findings. While this approach is designed to promote benign overfitting, it also presents a promising strategy for constructing effective
quantum kernels.

\begin{acknowledgements}
     H.K. was supported in part by the French National Research Agency under grant ANR-19-CE23-0011. J.T. acknowledges financial support from the QuanTEdu-France program under grant ANR-22-CMAS-0001. J.T. also thanks Kais Hariz and Ugo Nzongani for helpful discussions.
\end{acknowledgements}

\bibliography{biblio}

\newpage

\onecolumn

\title{Benign Overfitting with Quantum Kernels\\(Supplementary Material)}
\maketitle

\appendix

\section{Quantum Encoding for Periodic Translation-Invariant Quantum Kernels}
\label{app:quantum_encoding_periodic}

This appendix formalizes a class of quantum feature maps inducing periodic,
translation-invariant kernels. These quantum kernels yield interpolating estimators that are statistically consistent (and thus benign overfitting) as described in Theorem~\ref{thm:per-quantum-benign} and shown in Appendices~\ref{app:setting} to~\ref{app:proof}.

\subsection{Illustrative Example: Angle Encoding}
\label{subsec:angleencoding}
Consider $d$ qubits initialized in $\rho_0=(\ket{0}\bra{0})^{\otimes d}\!$.
Given $x=(x_1,\dots,x_d)\in\mathbb{R}^d$, define single-qubit rotations over the $\sigma_X$-Pauli matrix on the $j$th-qubits
\[
U_j(x_j)=\exp\!\Big(-i\frac{x_j}{2}\sigma_X^{(j)}\Big)=R^{(j)}_X(x_j),
\]
and the encoding unitary
\[
U_{\cos^2}(x)=\prod_{j=1}^d U_j(x_j).
\]
The corresponding quantum feature map, expressed as a density matrix, is
\[
\rho_{\cos^2}(x)=U_{\cos^2}(x)\rho_0 U_{\cos^2}(x)^\dagger.
\]

The induced quantum kernel
\[
k_{\cos^2}(x,z)=\mathrm{Tr}\!\big[\rho_{\cos^2}(x)\rho_{\cos^2}(z)\big]
\]
factorizes and satisfies
\[
k_{\cos^2}(x,z)
=
\prod_{j=1}^d
\cos^2\!\Big(\frac{x_j-z_j}{2}\Big).
\]
For the details of the computations, see~\cite{canatar2023bandwidth}. This quantum kernel is translation invariant and $2\pi$-periodic in each coordinate.

Let $\tau=(\tau_1,\dots,\tau_d)\in(\mathbb{R}^+_*)^d$. Replacing $x_j$ by
$\frac{2\pi x_j}{\tau_j}$ yields the kernel
\[
k(x,z)
=
\prod_{j=1}^d
\cos^2\!\Big(\frac{\pi(x_j-z_j)}{\tau_j}\Big),
\]
which is periodic with period $\tau_j$ in the $j$-th coordinate. The spike set $Z$ is 
\begin{align*}
    Z&:= \{(x,z) \in \mathcal X^2 : k(x,z) = 1 \}
    = \{(x,z) \in \mathcal X^2 : x-z \in \tau \odot \mathbb Z ^d \},
\end{align*}
where $\odot$ is the element-wise product. The spike set is formed of the input data points such that the pairwise difference lies  in the lattice parametrized by $\tau$. 

\subsection{Hamiltonian Encoding}
\label{subsec:hamiltonian}
The above construction admits a Hamiltonian representation
\[
U_{\cos^2}(x)
=
\exp\!\Big(-i\sum_{j=1}^d H_j x_j\Big),
\qquad \text{with }
H_j=\tfrac12\,\sigma_X^{(j)}.
\]
Since the $H_i$ act nontrivially on different qubits, the $H_i$ commute, i.e. $[H_j,H_k]=0$. One has $U_{\cos^2}(x)U_{\cos^2}^\dagger(z)=U_{\cos^2}(x-z)$ which leads to translation invariance of the kernel.

To obtain a larger class of quantum kernels that have the same translation invariance and periodicity properties, we construct the unitary encoding from its Hamiltonian representation and the quantum feature map in the following way.  

Let
\[
U(x)=\exp\!\Big(-i\sum_{j=1}^d H_j x_j\Big),
\,\,
\rho(x)=U(x)\ket{0^m}\!\bra{0^m}U(x)^\dagger,
\]
where $\{H_j\}_{j=1}^d$ are Hermitian operators acting on $m$ qubits satisfying
\begin{itemize}
    \item[\textbf{(F1)}] $[H_j,H_k]=0$ for all $j,k$;
    \item[\textbf{(F2)}] $\ket{0^m}$ is not an eigenstate of $H_j$  for all $j$;
    \item[\textbf{(F3)}] For each $j$, there exists a minimal $2\pi \ge \tau_j>0$ such that
    $e^{-iH_j\tau_j}\ket{0^m}=e^{-i\theta_j}\ket{0^m}$ for some $\theta_j\in\mathbb{R}$.
\end{itemize}

Under \textbf{(F1)}, the induced kernel satisfies
\[
k(x,z)=\mathrm{Tr}\!\big[\rho(x)\rho(z)\big]
=
\big|\braket{0^m|U(x-z)|0^m}\big|^2
=
\kappa(x-z),
\]
and is therefore translation invariant.

Assumption \textbf{(F2)} ensures that the kernel function does not depend trivially on each coordinate. For example if $H_j\ket{0^m} = \theta_j \ket{0^m}$ then $\exp(-iH_j)\ket{0^m} = \exp(-i\theta_j) \ket{0^m}$. We have $|\braket{0^m|\exp\big(-iH_j(x_j-z_j)\big)|0^m}|^2 = |\exp(-i\theta_j(x_j-z_j))|^2 = 1$ which is independent of the $j$-th coordinates of $x-z$. 

If we add assumption \textbf{(F3)}, we have:
\[
\kappa(x-z)=1
\quad\Longleftrightarrow\quad
x_j-z_j\in\tau_j\mathbb{Z}
\quad
\forall j\in\{1,\dots,d\}.
\]

Hence the kernel is periodic with period vector
$\tau=(\tau_1,\dots,\tau_d)$ and its spike set is parametrized by a lattice in $\mathbb{R}^d$ determined by the hamiltonians $\{H_j\}$.

\begin{remark}
 A simple implementation is obtained by choosing the Hamiltonians $H_j$ as local operators acting on distinct qubits, so that (\textbf{F1}) holds automatically. More generally, since commuting Hamiltonians share a common eigenbasis, one can construct efficiently implementable examples by combining an efficiently implementable unitary change of basis with a data-parametrized diagonal operator, which can be implemented efficiently using techniques from~\citet{zylberman2025efficient}.
\end{remark}

\section{Setting and Notations}
\label{app:setting}

\subsection{Form of the Local Kernel}
Let
\[
\ket{\psi}:\mathcal X\subset\mathbb{R}^d\to\mathbb{C}^{2^m},
\qquad
\ket{\psi(x)}=U(x)\ket{0^m},
\]
be a quantum feature map with $\|\ket{\psi(x)}\|_2=1$ for all $x\in\mathcal X$.
Define the associated density matrix
\[
\rho(x)=\ket{\psi(x)}\bra{\psi(x)},
\quad
\|\rho(x)\|_{\mathrm{HS}}:= \sqrt{\mathrm{Tr}[\rho(x)\rho^\dagger(x)]}=1.
\]
The induced quantum kernel is
\begin{equation}
\label{def:quantum_kernel}
k(x,z)
=
\mathrm{Tr}\!\bigl[\rho(x)\rho(z)\bigr]
=
\bigl|\langle\psi(x)\mid\psi(z)\rangle\bigr|^2 .
\end{equation}
Note that $0\le k(x,z)\le 1, \forall (x,z)$.

Define the spike set as
\begin{equation}
\label{def:Z}
Z
:=
\{(x,z)\in\mathcal X^2 : k(x,z)=1\}, 
\end{equation}

and for $\delta>0$ its $\delta$-neighborhood
\begin{equation} \label{def:Sdelta}
    S(\delta)
:=
\Bigl\{(x,z)\in\mathcal X^2 :
\inf_{(x',z')\in Z}\|(x,z)-(x',z')\|_2\le\delta
\Bigr\}.
\end{equation}

We restrict attention to periodic translation-invariant kernels satisfying:
\begin{equation}
\label{hyp:periodicity}\tag{H1}
\exists\,\tau\in(\mathbb{R}^+_*)^d
\text{ s.t.}\quad
k(x,z)=1
\;\Longleftrightarrow\;
x-z\in\tau\odot\mathbb{Z}^d .
\end{equation}
Note that the kernels described in Appendix~\ref{app:quantum_encoding_periodic} all have this property.
Under \emph{(H1)}, the spike set admits the explicit characterization
\[
Z=\{(x,z)\in\mathcal X^2 : x-z\in\tau\odot\mathbb{Z}^d\}
\]
and we have: $0\le k(x,z)<1, \forall (x,z)\notin Z$.

\subsection{Data Setting}
\label{subsec:data_setting}

Let $\mathcal X\subset\mathbb{R}^d$ be bounded.
Let $\mu$ be a probability measure on $\mathcal X$ admitting a density
$g\in L^1(\mathbb{R}^d)\cap L^p(\mathbb{R}^d)$ for some $p>1$ (possibly $p =+\infty$), with full support:
\[
\forall x\in\mathcal X,\ \forall t>0,\quad \mu(B(x,t))>0.
\]

Let $\{(x_i,y_i)\}_{i=1}^n$ be i.i.d.\ samples generated as
\begin{equation}
\label{eq:data_model}
x_i\sim\mu,
\qquad
y_i=f^\ast(x_i)+\varepsilon_i,
\end{equation}
where $\mathbb{E}[\varepsilon_i\mid x_i]=0$,
$\mathbb{E}[\varepsilon_i^2\mid x_i]=\sigma^2$, and
$f^\ast\in\mathcal H_k$ where $\mathcal H_k$ is the RKHS of the local kernel $k$.

Denote by
\[
X_n=\{x_1,\dots,x_n\},
\qquad
Y_n=(y_1,\dots,y_n)^\top.
\]

Define the minimal pairwise spike set distance as
\begin{equation}
\label{eq:delta_def}
\Delta(X_n)
=
\min_{1\le i<j\le n}
\;
\inf_{(x',z')\in Z}
\|(x_i,x_j)-(x',z')\|_2 .
\end{equation}

\subsection{Kernel Regressors}
Let
\[
K=(k(x_i,x_j))_{i,j},
\qquad
K^{\odot q}=(k(x_i,x_j)^q)_{i,j}.
\]
For $x\in\mathcal X$, define
\begin{align*}
    k(x,X_n)&=(k(x,x_1),\dots,k(x,x_n))^\top, \\
    k^{\odot q}(x,X_n)&=(k(x,x_1)^q,\dots,k(x,x_n)^q)^\top.
\end{align*}

The $\rho$-ridge regressor obtained  with the kernel $k$ is
\[
\hat f_{\rho,n}(x)
=
k(x,X_n)^\top (K+\rho I_n)^{-1}Y_n.
\]

For $q\in\mathbb{N}^*$ and $\rho>0$, using the separable global encoding in Fig.~\ref{fig:circuit1}, obtain a Local-Global kernel of the form
\[
k_{\rho,q}(x,z)=k(x,z)+\rho\,k(x,z)^q,
\]
with the Gram matrix $K_{\rho,q}=K+\rho K^{\odot q}$.

The ridgeless regressor obtained with the kernel $k_{\rho,q}$  is
\[
\hat f_{\rho,q,n}(x)
=
k_{\rho,q}(x,X_n)^\top K_{\rho,q}^+ Y_n,
\]
where $(\cdot)^+$ denotes the Moore-Penrose pseudoinverse.

\section{Auxiliary Results on the Spike Set}
\label{app:auxres}

\begin{lemma}
\label{lem:structureC}
Assume~\eqref{hyp:periodicity} and $\mathcal X\subset\mathbb R^d$ bounded.
Then there exists a finite set
$\{a_i\}_{i=1}^{N_{\mathcal X}}\subset\mathbb Z^d$, with $a_1=0$, such that
\[
Z
=
\bigcup_{i=1}^{N_{\mathcal X}}
\bigl\{(x,z)\in\mathcal X^2:\ x-z=\tau\odot a_i\bigr\}.
\]
\end{lemma}

\begin{proof}
By~\eqref{hyp:periodicity},
\begin{align}
    Z=\{(x,z)\in\mathcal X^2:\ x-z=\tau \odot \mathbb Z^d\}=\bigcup_{m\in\mathbb Z^d}
\{(x,z)\in\mathcal X^2:\ x-z=\tau\odot m\}.
\label{eq:infiniteunion}
\end{align}

Consider $m\in \mathbb Z^d$ such that  $\{(x,z)\in\mathcal X^2:\ x-z=\tau\odot m\}\not= \emptyset$. There exists $(x,z)\in \mathcal X^2$ such that $x-z=\tau\odot m$. Thus
\[
\|\tau\odot m\|_2\le 2\sup_{x\in\mathcal X}\|x\|_2
\]

Also $\|\tau\odot m\|_2\ge \tau_{\min}\|m\|_2$ with
$\tau_{\min}=\min_j\tau_j>0$ by~\eqref{hyp:periodicity}.

Since $\mathcal X$ is bounded, we conclude that $\|m\|_2 \le \tfrac{2\sup_{x\in\mathcal X}\|x\|_2}{\tau_{\min}}<+\infty$. So  there is only a finite number of  $m$ in $\mathbb Z^d$ yielding non-empty sets in~\eqref{eq:infiniteunion}.

The claim follows, noting that $m=0_d$ is always included since $(x,x) \in \mathcal X^2$ for all $x \in \mathcal X$.
\end{proof}

\begin{lemma}
\label{lem:structureS}
Assume~\eqref{hyp:periodicity} and $\mathcal X\subset\mathbb R^d$ bounded.
Then there exists a finite collection
$\{a_i\}_{i=1}^{N_{\mathcal X}}\subset\mathbb Z^d$, with $a_1=0$, such that
for every $\delta>0$,
\[
S(\delta)
\subset
\bigcup_{i=1}^{N_{\mathcal X}}
\Bigl\{(x,z)\in\mathcal X^2 :
\|x-z-\tau\odot a_i\|_2 \le \sqrt{2}\,\delta
\Bigr\}.
\]
\end{lemma}

\begin{proof}
By Lemma~\ref{lem:structureC}, the spike set admits a finite decomposition
\[
Z=\bigcup_{i=1}^{N_{\mathcal X}} Z_{a_i},
\qquad
Z_{a_i}:=\{(x,z)\in\mathcal X^2:\ x-z=\tau\odot a_i\}.
\]
Hence,
\[
S(\delta)
=
\bigcup_{i=1}^{N_{\mathcal X}}
\Bigl\{(x,z)\in\mathcal X^2:
\inf_{(x',z')\in Z_{a_i}}\|(x,z)-(x',z')\|_2\le\delta
\Bigr\}.
\]
For each $i$, define the set
\[
W_{a_i}:=\{(x',z')\in\mathbb R^{2d}:\ x'-z'=\tau\odot a_i\},
\]
so that $Z_{a_i}\subset W_{a_i}$ and
\[
\inf_{(x',z')\in Z_{a_i}}\|(x,z)-(x',z')\|_2
\ge
\inf_{(x',z')\in W_{a_i}}\|(x,z)-(x',z')\|_2.
\]
Note that
\[
\inf_{(x',z')\in W_{a_i}}\|(x,z)-(x',z')\|^2_2
=
\inf_{w\in\mathbb R^d}
\|(x,z)-(w,w-\tau\odot a_i)\|^2_2.
\]
The latter is  a strictly convex quadratic form in $w$, a direct computation yields
\[
\inf_{(x',z')\in W_{a_i}}\|(x,z)-(x',z')\|_2^2
=
\tfrac12\,\|x-z-\tau\odot a_i\|_2^2,
\]
where the minimizer is $(x',z') = (w^*,w^*-\tau \odot a_i)$ and
$
w^*=\tfrac12(x+z+\tau\odot a_i).
$
Therefore,
\[
\inf_{(x',z')\in Z_{a_i}}\|(x,z)-(x',z')\|_2\le\delta
\Rightarrow\;
\|x-z-\tau\odot a_i\|_2\le\sqrt{2}\,\delta,
\]
which concludes the proof.

\end{proof}

\begin{lemma}
\label{lem:Mn}
Assume $\mathcal X\subset\mathbb R^d$ is bounded.
Let $x_1,x_2\sim\mu$ i.i.d., where $\mu$ admits a density
$g$ fully supported on $\mathcal X$ with $g\in L^1(\mathbb R^d)\cap L^p(\mathbb R^d)$, for $p>1$. Let
$p'$ satisfy $\tfrac{1}{p}+\tfrac{1}{p'}=1$.
Then
\[
\forall z\in\mathcal X, \qquad 
\mu\bigl(\{x:(x,z)\in S(\delta)\}\bigr)
=
O\big(\delta^{\tfrac{d}{p'}}\big).
\]

\end{lemma}

\begin{proof}
Fix $z\in\mathcal X$. By Lemma~\ref{lem:structureS},
\begin{align*}
\{x:(x,z)\in S(\delta)\}\subset 
\bigcup_{i=1 }^{N_{\mathcal X}}\bigl\{x\in\mathcal X:\ \|x-z-\tau\odot a_i\|_2\le\sqrt{2}\,\delta\bigr\}
=\bigcup_{i=1 }^{N_{\mathcal X}}\left(B(z+\tau\odot a_i,\sqrt{2}\delta)
\cap \mathcal X\right).
\end{align*}

Therefore,
\begin{align*}
\mu\big(\{x:(x,z)\in S(\delta)\}\big)
&=
\int_{\mathcal X}
\ind_{\{(x,z)\in S(\delta)\}}(x)\,g(x)\,dx
\le
\sum_{i=1}^{N_{\mathcal X}}
\int_{\mathcal X}
\ind_{B(z+\tau\odot a_i,\sqrt{2}\delta)}(x)\,g(x)\,dx.
\end{align*}

Applying Hölder’s inequality with exponents $(p,p')$ gives
\begin{align*}
\int_{\mathcal X}
\ind_{B(z+\tau\odot a_i,\sqrt{2}\delta)}(x)\,g(x)\,dx
\le&
\|g\|_{L^p}\,
\|\ind_{B(z+\tau\odot a_i,\sqrt{2}\delta)}\|_{L^{p'}}
=
\|g\|_{L^p}\,
\bigl(v_d(\sqrt{2}\delta)^d\bigr)^{1/p'},
\end{align*}
where $v_d$ denotes the volume of the unit ball in $\mathbb R^d$.

Combining the above bounds concludes the proof:
\[
\mu\big(\{x:(x,z)\in S(\delta)\}\big)
\le
 N_{\mathcal X}\|g\|_{L^p}  (v_d\sqrt 2)^{\tfrac{d}{p'}}
\delta^{\frac{d}{p'}}.
\]

\end{proof}

\begin{lemma}
\label{lem:pn}
Under the assumptions of Lemma~\ref{lem:Mn},we have
\[
p(\delta)
:=
(\mu\times\mu)\big(S(\delta)\big)
=
O\big(\delta^{\tfrac{d}{p'}}\big).
\]
\end{lemma}

\begin{proof}
    \begin{align*}
    p(\delta)&:=(\mu \times \mu)(\{(x,z)\in S(\delta)\})\\
    &= \int_{\mathcal{X}^2} \ind_{S(\delta)}(x,z)g(x)g(z) dxdz \\
    &=\int_{\mathcal{X}}\Big( \mu\big(\{x:(x,z)\in S(\delta)\}\big)\Big)g(z) dz \\
    & \le \int_{\mathcal{X}} N_{\mathcal X}\|g\|_{L^p}  (v_d\sqrt 2)^{\tfrac{d}{p'}}
\delta^{\frac{d}{p'}} g(z) dz \qquad \text{(using  Lemma~\ref{lem:Mn})} \\
& \le N_{\mathcal X}\|g\|_{L^p}  (v_d\sqrt 2)^{\tfrac{d}{p'}}\delta^{\frac{d}{p'}} ,
\end{align*}
which concludes the proof.
\end{proof}

\begin{lemma}
\label{lem:min-dist}
Assume the same conditions as in Lemma~\ref{lem:Mn}.
Fix $\alpha>0$. 
If the sequence $\{\delta_n\}_{n\in\N}$ verifies $\delta_n=O\left(n^{-(2+\alpha)\tfrac{p'}{d}}\right)$
and if for all $n$, $X_n=\{x_i\}_{i=1}^n$ are i.i.d.\ samples from $\mu$,
then
\[
\mathbb P\big(\Delta(X_n)>\delta_n\big)\xrightarrow[n\to\infty]{}1.
\]
where
        \[
        \Delta(X_n) = \min_{1\le i<j\le n} \inf_{(x',z')\in Z} \|(x_i,x_j)-(x',z')\|_2.
        \]
\end{lemma}

\begin{proof}
Noticing that
\begin{align*}
    \Delta(X_n)\le \delta_n 
    &\Leftrightarrow \exists\ (i,j)\in\{1,..,n\}^2, {i<j}, \text{ s.t.} \ \inf_{(x',z')\in Z} \|(x_i,x_j)-(x',z')\|_2 \le \delta_n\\
    &\Leftrightarrow \exists\ (i,j)\in\{1,..,n\}^2,  {i<j}, \text{ s.t.}  \ (x_i,x_j)\in S(\delta_n).
\end{align*}
We have :
$\forall(x_1,..,x_n), \ \ind_{\{\Delta(X_n)\le \delta_n \}}(x_1,..,x_n)\le \sum_{i=1}^n\sum_{j=i+1}^n \ind_{S(\delta_n)}(x_i,x_j)$, therefore
\begin{align*}
    \mathbb P(\Delta(X_n)\le \delta_n )
    & \le \int_{\X^n}\left(\sum_{i=1}^n\sum_{j=i+1}^n \ind_{S(\delta_n)}(x_i,x_j) \right)g(x_1)..g(x_n)dx_1..dx_n \\
    & \le \sum_{i=1}^n\sum_{j=i+1}^n \int_{\X^n}\ind_{S(\delta_n)}(x_i,x_j) g(x_1)..g(x_n)dx_1..dx_n \\
    & \le \sum_{i=1}^n\sum_{j=i+1}^n p(\delta_n) \\
    & \le \tfrac{n(n-1)}{2} p(\delta_n)\\
    & \le C\tfrac{n(n-1)}{2} (\delta_n)^{\frac{d}{p'}},
\end{align*}
where the last line follows from Lemma~\ref{lem:pn}. 
By assumption $\delta_n \le C'\left(n^{-(2+\alpha)\tfrac{p'}{d}}\right) $ 
so that 
$C\tfrac{n(n-1)}{2} (\delta_n)^{\frac{d}{p'}}
\le \tfrac{C {C'}^{\frac{d}{p'}}}{2} n^{-\alpha}$.
We conclude that  
$\lim_{n\to +\infty}\mathbb P(\Delta(X_n)\le \delta_n ) =0$ and
 $\lim_{n\to +\infty}\mathbb P(\Delta(X_n)> \delta_n ) =1$.
 \end{proof}

\section{Proof of the Main Theorem}
\label{app:proof}

\begin{theorem}[Benign overfitting for periodic quantum kernels]
        \label{thm:per-quantum-benign}
         Let $k$ be a quantum fidelity kernel satisfying~\eqref{hyp:periodicity}, let the data satisfy the hypotheses in Appendix~\ref{subsec:data_setting}. 
         Write $\beta = \frac{d}{p'}$ with $\frac{1}{p'}+\frac{1}{p}=1$.
Choose $\alpha>0$ and $0<\alpha_0<1$. 
        
        Then choosing sequences $\{\delta_n\}_{n\ge1}$, $\{q_n\}_{n\ge1}$ and
        $\{\rho_n\}_{n\ge1}$ satisfying:
        \[
        \delta_n = o\left(n^{-\frac{2+\alpha}{\beta}}\right),
        \quad
        \quad \rho_n n^{\alpha_0} \xrightarrow[n\to\infty]{}0, \quad 
        \rho_n n \xrightarrow[n\to\infty]{} +\infty,
        \]
        and
        \begin{equation*}
        q_n \ge \frac{c_\alpha \log n}{-\log m(\delta_n)},
        \qquad     \text{with}\qquad 
        c_\alpha>\tfrac{6+\alpha}{2},
        \end{equation*}
        where
        \[
        m(\delta)\eqdef\sup_{(x,z)\in S(\delta)^c} k(x,z),\quad \forall \delta>0.
        \]

        \[
        \text{Write}\qquad
        \Delta(X_n) = \min_{1\le i<j\le n} \inf_{(x',z')\in Z} \|(x_i,x_j)-(x',z')\|_2
        \qquad\text{and}\qquad
        P_n = \mathbb P( \Delta(X_n) < \delta_n).
        \]
        Then $P_n \xrightarrow[n\to \infty]{}0$ and with probability at least $(1-O(n^{-\alpha_0}))(1-O(n^{-\alpha}))(1-P_n)$, the ridgeless estimator $\hat f_{\rho_n,q_n,n}$, built on the kernel $k_{\rho_n,q_n}$, 
        satisfies
        \begin{equation}
        \label{eq:error_dec0}
        \E_{x}\!\left[
        \big|\hat f_{\rho_n,q_n,n}(x)-f^*(x)\big|^2
        \right] \le B_{0,n}+B_{1,n}+B_{2,n}\;.
        \end{equation}
        
        with 
        \begin{align}
        B_{0,n}& = 2\sigma^2 4^m n^{\alpha_0-1}\Big(1+\frac{1}{\rho_n n}\Big)  \quad +\quad 2\|f^*\|_{\mathcal H_k}^2 \rho_n n^{\alpha_0}\Big(1+\frac{1}{\rho_n n}\Big)^2
        ,\label{eq:B0n}\\
        B_{1,n}&=4C_y\,
           \frac{n^{4+\alpha}m(\delta_n)^{2q_n}}
                {\rho_n^2\big(1-nm(\delta_n)^{q_n}\big)^2},  &\quad \text{where}\quad C_y = \mathbb{E}_x[f^*(x)^2]+\sigma^2,\label{eq:B1n}\\
        B_{2,n}& = 4C_y\,
           \frac{n^{2+\alpha}\left(C_\mu\delta_n^\beta +m(\delta_n)^{2q_n}\right)}
                {\big(1-nm(\delta_n)^{q_n}\big)^2} ,
                 &\quad\text{where}\quad
     C_\mu= N_{\mathcal X}\|g\|_{L^p}  (v_d\sqrt 2)^{\tfrac{d}{p'}}
                \label{eq:B2n}.
        \end{align}
        
        Furthermore $B_{i,n}\xrightarrow[n \to \infty]{} 0 $ for $i=0,1,2$.
        
        Hence, the sequence of estimators $\{\hat f_{\rho_n,q_n,n}\}_n$ is consistent in probability.
    \end{theorem}

\begin{proof}[Proof of Theorem~\ref{thm:per-quantum-benign}]

We follow the same error decomposition as in \cite{haas2023mind}.
\begin{equation}
\label{eq:decerror}
\E_{x}\!\left[
        \big|\hat f_{\rho,q,n}(x)-f^*(x)\big|^2
        \right]
        \le 2\E_{x}\!\left[
        \big|\hat f_{\rho,n}(x)-f^*(x)\big|^2
        \right]
        +2\E_{x}\!\left[
        \big|\hat f_{\rho,q,n}(x)-\hat f_{\rho,n}(x)\big|^2
        \right]
        .
\end{equation}
and
\begin{equation*}
\E_{x}\!\left[
        \big|\hat f_{\rho,q,n}(x)-\hat f_{\rho,n}(x)\big|^2
        \right]
        \le 2 \E_{x}\!\left[
        \big|k(x,X_n)^\top\big[(K+\rho I_n)^{-1}-K_{\rho,q}^+\big]Y_n\big|^2
        \right]
        +2\rho^2\E_{x}\!\left[
        \big|k^{\odot q}(x,X_n)^\top K_{\rho,q}^+Y_n\big|^2
        \right].
\end{equation*}

Applying Cauchy-Schwarz' inequality and operator norm inequality gives
\begin{align}
\E_{x}\!\left[
        \big|\hat f_{\rho,q,n}-\hat f_{\rho,n}\big|^2
        \right]
        \le 2\,\E_x\big[\norm{k(x,X_n)}_2^2\big]\,
     \big\|K_{\rho,q}^+-(K+\rho I_n)^{-1}\big\|_{\mathrm{op}}^2\,
     \norm{Y_n}_2^2 
 +2\rho^2\,\E_x\big[\norm{k^{\odot q}(x,X_n)}_2^2\big]\,
     \|K_{\rho,q}^+\|_{\mathrm{op}}^2\,
     \norm{Y_n}_2^2
     \label{eq:matineq}
\end{align}

Combining~\eqref{eq:decerror} and~\eqref{eq:matineq}, we obtain
\begin{equation}
\label{eq:error_dec}
\E_{x}\!\left[
        \big|\hat f_{\rho,q,n}(x)-f^*(x)\big|^2
        \right]\le A_{0,n}+A_{1,n}+A_{2,n},
\end{equation}
where
\begin{align}
\label{eq:Aterms}
A_{0,n}&\eqdef2\E_{x}\!\left[
        \big|\hat f_{\rho,n}(x)-f^*(x)\big|^2
        \right],\\
A_{1,n}&\eqdef4\,\E_x\big[\norm{k(x,X_n)}_2^2\big]\,
          \big\|K_{\rho,q}^+-(K+\rho I_n)^{-1}\big\|_{\mathrm{op}}^2\,
          \norm{Y_n}_2^2,\\
A_{2,n}&\eqdef4\rho^2\,\E_x\big[\norm{k^{\odot q}(x,X_n)}_2^2\big]\,
          \|K_{\rho,q}^+\|_{\mathrm{op}}^2\,
          \norm{Y_n}_2^2.
\end{align}
To show the consistency in probability a sufficient condition is to obtain $A_{i,n} \xrightarrow[n \to \infty]{}0$ in probability for $i=0,1,2$.

\paragraph{Bounding $A_{0,n}$\\}

By construction, the quantum feature map of $k$ in Eq~\eqref{def:quantum_kernel} takes values in the Hilbert space of linear operators on $(\mathbb C^2)^{\otimes m}$ endowed with the Hilbert-Schmidt inner product. Since $\rho(x)$ is a rank-one density matrix, we have
\[
\forall x \in \mathcal X,\quad \|\rho(x)\|_{\mathrm{HS}}^2 = \operatorname{Tr}\!\big[\rho(x)\rho^\dagger(x)\big] = \operatorname{Tr}\!\big[\rho^2(x)\big] = 1
\]
so the feature map is uniformly bounded by 1. Moreover, this Hilbert space has finite dimension $p \le 4^m$.

So, under the quantum feature map in Eq~\eqref{def:quantum_kernel} and data hypothesis in Appendix~\ref{subsec:data_setting}, we can apply the Proposition~7.6 of~\cite{bach2024learning} with regularization $\lambda=\rho_n,R=1$ in the finite dimensional case 

\begin{equation}
\E_{(X_n,Y_n)}\left[\E_x \left[|\hat f_{\rho_n,n}(x)-f^*(x)|^2\right]\right]
\le
\frac{\sigma^2 p}{n}\Big(1+\frac{1}{\rho_n n}\Big)
+
\rho_n\Big(1+\frac{1}{\rho_n n}\Big)^2
\|f^*\|_{\mathcal H_k}^2.
\end{equation}

When choosing a regularization sequence satisfying
\[
\rho_n\to 0
\qquad\text{and}\qquad
\rho_n n\to\infty,
\]
both terms on the right-hand side vanish:
\[
\E_{(X_n,Y_n)}\left[\E_x \left[|\hat f_{\rho_n,n}(x)-f^*(x)|^2\right]\right]
\xrightarrow[n\to\infty]{}
0.
\]

Let
\[
C_n(\epsilon) =\big\{ (X_n,Y_n) \in (\mathcal X \times \mathbb R)^n:  \E_x \left[|\hat f_{\rho_n,n}(x)-f^*(x)|^2\right] \ge \epsilon\big\}
\]
By Markov's inequality, for any fixed $\alpha_0>0$,
\begin{align*}
\mathbb P \Big( C_n\big(  \E_x \left[|\hat f_{\rho_n,n}(x)-f^*(x)|^2\right]\,n^{\alpha_0} \big) \Big)
\le n^{-\alpha_0}. 
\end{align*}

Hence, with probability at least $(1-n^{-\alpha_0})$,
\begin{equation}
\label{eq:A0n}
A_{0,n}\le 
\frac{2\sigma^2 p n^{\alpha_0}}{n}\Big(1+\frac{1}{\rho_n n}\Big)
+
2n^{\alpha_0}\rho_n\Big(1+\frac{1}{\rho_n n}\Big)^2
\|f^*\|_{\mathcal H_k}^2 = B_{0,n}.
\end{equation}

Then choosing $0<\alpha_0<1$ and $\{\rho_n\}_n$ such that
\[
  n^{\alpha_0}\rho_n\xrightarrow[n\to \infty ]{}0,\quad  n\rho_n\xrightarrow[n\to \infty ]{}+\infty 
\]
guarantees that $B_{0,n} \xrightarrow[n \to \infty]{} 0$ and $A_{0,n} \xrightarrow[n \to \infty]{} 0$ with probability at least $(1-n^{-\alpha_0})$.

\paragraph{Bounding $A_{2,n}$\\}

We have $A_{2,n}\eqdef4\rho^2\,\E_x\big[\norm{k^{\odot q}(x,X_n)}_2^2\big]\,
          \|K_{\rho,q}^+\|_{\mathrm{op}}^2\,
          \norm{Y_n}_2^2.$  We bound each term.

\subparagraph{Bounding $\mathbb{E}_x\!\big[\|k^{\odot q_n}(x,X_n)\|_2^2\big]$\\}

For each $i$:
\begin{align*}
    k^{2q_n}(x,x_i) &= k^{2q_n}(x,x_i)\left( \ind_{\{(x,x_i)\in S(\delta_n)\}}+\ind_{\{(x,x_i)\in S^c(\delta_n)}\right)  \\
    &\le \ind_{\{(x,x_i)\in S(\delta_n)\}}+m(\delta_n)^{2q_n}\ind_{\{(x,x_i)\in S^c(\delta_n)\}} \\
    & \le \ind_{\{(x,x_i)\in S(\delta_n)\}}+m(\delta_n)^{2q_n}
\end{align*}

Then summing over $i$ and taking the expectation gives
\begin{align*}
    \E_x \big[\norm{k^{\odot q_n}(x,X_n)}_2^2\big]& = \sum_{i=1}^n\E_x[k^{2q_n}(x,x_i)]
\le\sum_{i=1}^n\mu\big(\{x:(x,x_i)\in S(\delta_n)\}\big)
   +n m(\delta_n)^{2q_n} .
\end{align*}

Then using Lemma~\ref{lem:Mn}, we obtain 
\begin{equation*}
\E_x\big[\norm{k^{\odot q_n}(x,X_n)}_2^2\big]\le n(C_\mu \delta_n^\beta + m(\delta_n)^{2q_n}) \quad\text{where}\quad
     C_\mu= N_{\mathcal X}\|g\|_{L^p}  (v_d\sqrt 2)^{\tfrac{d}{p'}}.
\end{equation*}

\subparagraph{Bounding $\|K_{\rho_n,q_n}^+\|_{op}$\\}

Since $\delta_n = O\left(n^{-\frac{2+\alpha}{\beta}}\right)$, we can use Lemma~\ref{lem:min-dist}: with probability $(1-P_n)$, $\Delta(X_n) \ge \delta_n$ i.e. $(x_i,x_j)\notin S(\delta_n)$ for all $i\neq j$.
 
Write $K_{\rho_n,q_n}=K+\rho_n K^{\odot q_n}=K+\rho_n I_n+\rho_n E(q_n)$.
By Weyl's inequality,
\[
\lambda_{\min}(K_{\rho_n,q_n})
\ge\lambda_{\min}(K+\rho_n I_n)+\rho_n\,\lambda_{\min}(E(q_n)).
\]

For $i\neq j$, $|E(q_n)_{ij}|=k^{q_n}(x_i,x_j)\le m(\delta_n)^{q_n}$
and $E(q_n)_{ii}=0$. The Gershgorin circle theorem allows to bound all the eigenvalues of $E(q_n)$
\begin{align*}
    |\lambda_\ell(E(q_n))|&\le(n-1)m(\delta_n)^{q_n}.
\end{align*}
Hence $\|E(q_n)\|_{\mathrm{op}}\le nm(\delta_n)^{q_n}$ and $\lambda_{\min}(E(q_n))\ge -\|E(q_n)\|_{\mathrm{op}}\ge -nm(\delta_n)^{q_n}$ and
\[
\lambda_{\min}(K_{\rho_n,q_n})\ge\rho_n-\rho_n nm(\delta_n)^{q_n}
      =\rho_n\big(1-nm(\delta_n)^{q_n}\big).
\]

Notice that $0\le m(\delta_n)<1$. So since  $q_n \ge -\frac{\log n}{\log m(\delta_n)}$ then  $0 \le nm(\delta_n)^{q_n} < 1$. This implies that
 $\lambda_{\min}(K_{\rho_n,q_n})>0$ and $K_{\rho_n,q_n}^+=K_{\rho_n,q_n}^{-1}$. Consequently
\begin{equation*}
\|K_{\rho_n,q_n}^+\|_{\mathrm{op}}^2\le\frac{1}{\rho_n^2\big(1-nm(\delta_n)^{q_n}\big)^2}, \quad\text{ with probability }(1-P_n). 
\end{equation*}

\subparagraph{Bounding $\|Y_n\|^2_2$\\}
As in the proof of Theorem G.8 of \cite{haas2023mind}:
\begin{align*}
    \E\!\big[\norm{Y_n}_2^2\big]
=\sum_{i=1}^n\E[y_i^2]
=n\big(\E_x[f^*(x)^2]+\sigma^2\big)
 = n C_y .
\end{align*}

By Markov's inequality, for any fixed $\alpha>0$,
\[
\proba\Big(\norm{Y_n}_2^2\ge \E[\norm{Y_n}_2^2]\,n^{\alpha}\Big)
\le\frac{\E[\norm{Y_n}_2^2]}{\E[\norm{Y_n}_2^2]\,n^{\alpha}}=n^{-\alpha}.
\]
Hence, with probability at least $(1-n^{-\alpha})$,
$ \norm{Y_n}_2^2\le C_y\,n^{1+\alpha}.$

\subparagraph{Final bound for  $A_{2,n}$\\}
Collecting the three previous bounds yields with probability  at least $(1-n^{-\alpha})(1-P_n)$
\begin{align}
A_{2,n}
&\le 4\rho_n^2n\big(C_\mu\delta_n^\beta+m(\delta_n)^{2q_n}\big)\,
    \frac{ C_y n^{1+\alpha} }{\rho_n^2\big(1-nm(\delta_n)^{q_n}\big)^2}\,
   =  4 C_y\,
    \frac{ n^{2+\alpha} \big(C_\mu\delta_n^\beta+m(\delta_n)^{2q_n}\big)}{\big(1-nm(\delta_n)^{q_n}\big)^2}\,
   = B_{2,n}.
   \label{eq:A2n}
\end{align}

\paragraph{Bounding $A_{1,n}$\\}
\[
A_{1,n}\eqdef4\,\E_x\big[\norm{k(x,X_n)}_2^2\big]\,
          \big\|K_{\rho,q}^+-(K+\rho I_n)^{-1}\big\|_{\mathrm{op}}^2\,
          \norm{Y_n}_2^2.
\]
As before, we bound each term composing $A_{1,n}$. 

Since $k(x,x_i)^2\le 1$,
$\mathbb{E}_{x}\!\left[\|k(x,X_n)\|^2\right] = \mathbb{E}_{x}\!\left[\sum_{i=1}^n{k(x,x_i)^2}\right] \le n.
$

Using Lemma G.8 of~\cite{haas2023mind} with $A=K+\rho_n I_n$, $B=K_{\rho_n,q_n}$, we obtain
\begin{align*}
    \big\|K_{\rho_n,q_n}^+-(K+\rho_n I_n)^{-1}\big\|_{\mathrm{op}}
&\le\big\|(K+\rho_n I_n)^{-1}\big\|_{\mathrm{op}}\,
   \frac{\|E(q_n)\|_{\mathrm{op}}}{1-\|E(q_n)\|_{\mathrm{op}}} 
   \le\frac{1}{\rho_n}\,
   \frac{nm(\delta_n)^{q_n}}{(1-nm(\delta_n)^{q_n})}.
\end{align*}
Combining the previous bounds shows that with probability at least $(1-n^{-\alpha})(1-P_n)$
\begin{align}
A_{1,n}
&\le 4 n\,
    \frac{n^2m(\delta_n)^{2q_n}}{\rho_n^2\big(1-nm(\delta_n)^{q_n}\big)^2}\,
    C_y n^{1+\alpha} \nonumber\\
&=4C_y\,
   \frac{n^{4+\alpha}m(\delta_n)^{2q_n}}
        {n^2\rho_n^2\big(1-nm(\delta_n)^{q_n}\big)^2} = B_{1,n}.\label{eq:A1n} 
\end{align}

The bounds hold also with probability at least $(1-n^{-\alpha})(1-P_n)$.

\paragraph{Limits of $B_{1,n}$ and $B_{2,n}$\\}

To conclude the proof, it remains to show that  the hypotheses on $\{\delta_n\}_n$, $\{\rho_n\}_n$ and $\{q_n\}_n$ ensure that $B_{1,n} \xrightarrow[n\to \infty]{} 0$ and  $B_{2,n} \xrightarrow[n\to \infty]{} 0$.

Since $q_n \ge \frac{c_\alpha \log n}{-\log m(\delta_n)}$, we have
$ m(\delta_n)^{q_n}\leq n^{-C_\alpha}$ with $C_\alpha>\tfrac{6+\alpha}{2}$, thus 
\[
n^{2+\alpha}m(\delta_n)^{2q_n} \leq n^{2+\alpha-2C_\alpha}\leq n^{-4} \xrightarrow[n\to \infty]{} 0,
\]
\[
n^{4+\alpha}m(\delta_n)^{2q_n} \leq n^{4+\alpha-2C_\alpha}\leq n^{-2} \xrightarrow[n\to \infty]{} 0,
\]
and
\[
n m(\delta_n)^{q_n} \leq  n^{1-C_\alpha}\xrightarrow[n\to \infty]{} 0.
\]
By assumption $n\rho_n\xrightarrow[n\to\infty]{} +\infty$ and $n^{2+\alpha} \delta_n^\beta\xrightarrow[n\to\infty]{}0 $.
This gives all the ingredients to conclude 
that $B_{1,n}\xrightarrow[n\to\infty]{} 0$ and  $B_{2,n}\xrightarrow[n\to\infty]{} 0$.

\end{proof}

\section{Instantiation for the  Cosine Quantum Kernel}
\label{app:cos}

The purpose of this appendix is to connect the general periodic
result of Appendix~\ref{app:proof} with Theorem~\ref{thm:simplified-quantum-benign2}, which is stated specifically for the cosine-squared kernel $k(x,z)=\prod_{j=1}^d \cos^2\!\Big(\frac{x_j-z_j}{2}\Big)$.

Appendix~\ref{app:proof} establishes benign overfitting for periodic
translation-invariant kernels when $(q_n)$ grows at a rate which depends on $m(\delta)$. In contrast, Theorem~\ref{thm:simplified-quantum-benign2}
is stated with an explicit lower bound on $(q_n)$, which no longer involves $m(\delta)$. To bridge both results, we derive an explicit upper bound on $m(\delta)$ for the cosine kernel. Substituting this bound into the general condition of
Appendix~\ref{thm:per-quantum-benign} yields the explicit requirement on $(q_n)$
appearing in Theorem~\ref{thm:simplified-quantum-benign2}.

We now establish such a bound.

\begin{lemma}
\label{lem:cosbound}
Let $\delta > 0$ and
\[
k(x,z)
=
\prod_{j=1}^d
\cos^2\!\Big(\frac{x_j - z_j}{2}\Big).
\]
Then
\[
m(\delta)
\le
\exp\!\Big(
-\frac{2\delta^2}{\pi^2}
\Big).
\]
\end{lemma}

\begin{proof}
Since the kernel is $2\pi$-periodic in each coordinate,
Lemma~\ref{lem:structureS} implies that if $(x,z)\notin S(\delta)$,
then for some $a\in\mathbb Z^d$,
\[
h := x-z-2\pi a
\quad\text{satisfies}\quad
\|h\|_2>\sqrt{2}\,\delta
\quad\text{and}\quad
h\in[-\pi,\pi]^d.
\]

By convexity of $u \mapsto e^{-u}$ on $\mathbb{R}_+$, the function lies above
its tangent at $u=0$. Hence,
\[
\forall u \ge 0, \qquad 1 - u \le e^{-u}.
\]

Moreover, since $v \mapsto \sin v$ is concave on $[0,\pi]$, it lies above
the chord joining $(0,0)$ and $(\pi/2,1)$. Therefore,
\[
\forall v \in \left[0,\frac{\pi}{2}\right], 
\qquad 
\frac{2v}{\pi} \le \sin v.
\]

Let $t \in [-\pi,\pi]$. Setting $v = |t|/2 \in [0,\pi/2]$, we obtain
\[
\sin\!\left(\frac{|t|}{2}\right)
\ge
\frac{|t|}{\pi}.
\]

Squaring both sides yields
\[
\sin^2\!\left(\frac{t}{2}\right)
\ge
\frac{t^2}{\pi^2}.
\]

Using $\cos^2(t/2) = 1 - \sin^2(t/2)$, we deduce
\[
\cos^2(t/2)
\le
1 - \frac{t^2}{\pi^2}.
\]

Applying the inequality $1-u \le e^{-u}$ with $u = t^2/\pi^2$, we conclude
\[
\cos^2(t/2)
\le
\exp\!\left(-\frac{t^2}{\pi^2}\right).
\]

Hence
\[
k(x,z)
=
\prod_{j=1}^d \cos^2(h_j/2)
\le 
\exp\!\Big(-\frac{\|h\|_2^2}{\pi^2}\Big)
\le
\exp\!\Big(-\frac{2\delta^2}{\pi^2}\Big),
\]
which proves the claim.
\end{proof}

Appendix~\ref{app:proof} proves benign overfitting for periodic
translation-invariant kernels under the condition
\[
q_n \;\ge\; \frac{c_\alpha \log n}{-\log m(\delta_n)},
\qquad
c_\alpha > \tfrac{6+\alpha}{2}.
\]

For the cosine kernel, we have shown \ref{lem:cosbound} that
\[
m(\delta)\le \widetilde m(\delta)
:=\exp(-\frac{2\delta^2}{\pi^2}).
\]
Because $m(\delta_n)\le \widetilde m(\delta_n)<1$, we obtain
\[
-\log m(\delta_n)
\;\ge\;
-\log \widetilde m(\delta_n)
=
\frac{2\delta_n^2}{\pi^2}.
\]
Therefore
\[
\frac{c_\alpha \log n}{-\log m(\delta_n)}
\;\le\;
\frac{c_\alpha \pi^2 \log n}{\,2\delta_n^2},
\]
and it is sufficient to choose
\[
q_n \;\ge\; \frac{c_\alpha \pi^2 \log n}{\,2\delta_n^2},
\]

 Since $n^{-\xi} = o\!\left(n^{-\frac{2(2+\alpha)p}{d(p-1)}}\right)$ for $\xi > \frac{2(2+\alpha)p}{d(p-1)}$, we can set $\delta_n = \frac{\pi}{2}n^{\frac{-\xi}{2}} =o\!\left(n^{-\frac{(2+\alpha)p}{d(p-1)}}\right)$, giving $q_n \ge 2 c_\alpha n^\xi \ln n $, in accordance with Theorem \ref{thm:per-quantum-benign}. This yields the  condition on $(q_n)$ stated in Theorem~\ref{thm:simplified-quantum-benign2}.

\section{Additional Experimental Details}
\label{app:xp}

\subsection{Synthetic Experiments}

\begin{table}[h]
\centering
\caption{Summary of experimental settings. Dimension $d$; training set size $n$; test set size $n_{\text{test}}$; noise variance $\sigma^2$; kernel parameters $\rho,q,c$; target function.}
\begin{tabular}{lcccccccc}
\toprule
Experiment & $d$ & $n$ & $n_{\text{test}}$ & $\sigma^2$ & $\rho$ & $q$ & $c$ & Target \\
\midrule
a 
& $1$
& $50$
& $-$
& $0.5$
& $\sigma^2/\sqrt{n}$ 
& $\{2^p: p \in \{4,8,14,16\}\}$ 
& $\frac{3\pi}{4}$ 
& $f_1$~\eqref{eq:f1} 
\\
b1  
& $7$ 
& $500$ 
& $2000$
& $0.5$ 
& $\sigma^2/\sqrt{n}$ 
& $\{2^p: p \in \{2,4,6,8\}\}$ 
& $\frac{1}{\sqrt{d}}$
& $f_2$  ~\eqref{eq:target-general}~\eqref{eq:f2}
\\
b2 
& $7$ 
& $500$ 
& $2000$
& $7\cdot10^{-3}$ 
& $\sigma^2/\sqrt{n}$ 
& $\{2^p: p \in \{2,4,6,8\}\}$
& $\frac{1}{\sqrt{d}}$
& $f_3$  ~\eqref{eq:target-general}~\eqref{eq:f3}
\\
c 
& $7$ 
& $25$-$500$
& $2000$
& $0.5$ 
& $\sigma^2/\sqrt{n}$ 
& $\lceil 10 \log(n) n^{6/d}\rceil$ 
& $\frac{1}{\sqrt{d}}$
& $f_2$  ~\eqref{eq:target-general}~\eqref{eq:f2}
\\
\bottomrule
\end{tabular}
\label{tab:syntheticexperiments}
\end{table}

The synthetic experiments (a-c) illustrate the behavior of the Local-Global construction in controlled settings and complement the theoretical analysis.

\paragraph{General setting.}
Inputs are sampled i.i.d.\ from $\mathcal{U}([-1,1]^d)$ and outputs are generated as
\[
y_i = f^\star(x_i) + \sigma\varepsilon_i,
\qquad
\varepsilon_i \sim  \mathcal{N}(0,1).
\]
The base kernel is
\[
k_B(x,z) = \prod_{j=1}^d \cos^2\!\left(\frac{c(x_j - z_j)}{2}\right),
\]
and the Local-Global kernel is
\[
k_{LG}(x,z) = k_B(x,z) + \rho\, k_B(x,z)^q,
\]
with $\rho=\sigma^2/\sqrt{n}$ as in the table \ref{tab:syntheticexperiments}. For experiments (b) and (c), results are averaged over $10$ independent splits and evaluated using MSE on an independent test set.

\paragraph{Target functions.}
Experiment (a) uses the one-dimensional target
\begin{equation}
\label{eq:f1}
f_1(x)
=
\frac{2}{n_z}
\sum_{j=1}^{n_z}
k_B(x,z_j)
+ 1,
\qquad
z_j \overset{\mathrm{i.i.d.}}{\sim} \mathcal{U}([-1,1]),
\quad n_z=5.
\end{equation}

For experiments (b1), (b2), and (c), we generate targets  as:
\begin{equation}
\label{eq:target-general}
\tilde f(x; k^\star)
=
\sum_{j=1}^{n_z} \alpha_j\, k^\star(x,z_j)
-
\mathbb{E}_{x'}\!\left[
\sum_{j=1}^{n_z} \alpha_j\, k^\star(x',z_j)
\right],
\qquad
n_z=50, \qquad z_j \overset{\mathrm{i.i.d.}}{\sim} \mathcal{U}([-1,1]^d)
\end{equation}
where  $\alpha_j$ are i.i.d.\ Gaussian coefficients. We then rescale $\{\alpha_j\}$ so that, for $x\sim \mathcal{U}([-1,1]^d)$ we have $\mathbb{E}[\tilde f(x;k^\star)] = 0,\,\mathrm{Var}(\tilde f(x;k^\star)) = 1.$

In (b1) and (c), we take the base kernel:
\begin{equation}
\label{eq:f2}
k^\star(x,z)=k_B(x,z),
\end{equation}
and denote the resulting target by $f_2(x)=\tilde f(x;k_B) \in \mathcal{H}_{k_B}$.

In (b2), we take the Local-Global kernel
\begin{equation}
\label{eq:f3}
k^\star(x,z)
=
k_B(x,z)
+
\rho^\star k_B(x,z)^{q^\star},
\qquad
(\rho^\star,q^\star)=(5,8),
\end{equation}
and denote the resulting target by $f_3(x)=\tilde f(x;k^\star) \not \in  \mathcal{H}_{k_B} $.

\paragraph{Experiment (a).}
\label{par:a}
This one-dimensional setting provides a minimal illustration of the behavior of the Local-Global construction as $q$ varies, as depicted in Figure~\ref{fig:spsmreg}. We set the parameters as described in Table~\ref{tab:syntheticexperiments}. We compare ridge regression with $k_B$ ($\lambda=\rho$) and ridgeless regression with $k_{LG}$ for different values of $q$.
For numerical stability, a small regularization $\lambda_\varepsilon=10^{-14}$ is used when $q=2^4$; all other models are trained in the ridgeless regime. Predictions are evaluated on a dense grid over $[-1,1]$ for visualization. 

\paragraph{Experiment (b).}
\label{par:b}
We consider the setting described in Table~\ref{tab:syntheticexperiments} for experiments (b1) and (b2). We compare ridge regression with $k_B$ ($\lambda=\rho$), ridgeless regression with $k_B$ ($\lambda=0$), and ridgeless regression with $k_{LG}$ for the values of $q$ specified in the table. Results are averaged over $10$ independent splits.

\begin{figure}
    \centering
    \includegraphics[width=0.6\linewidth]{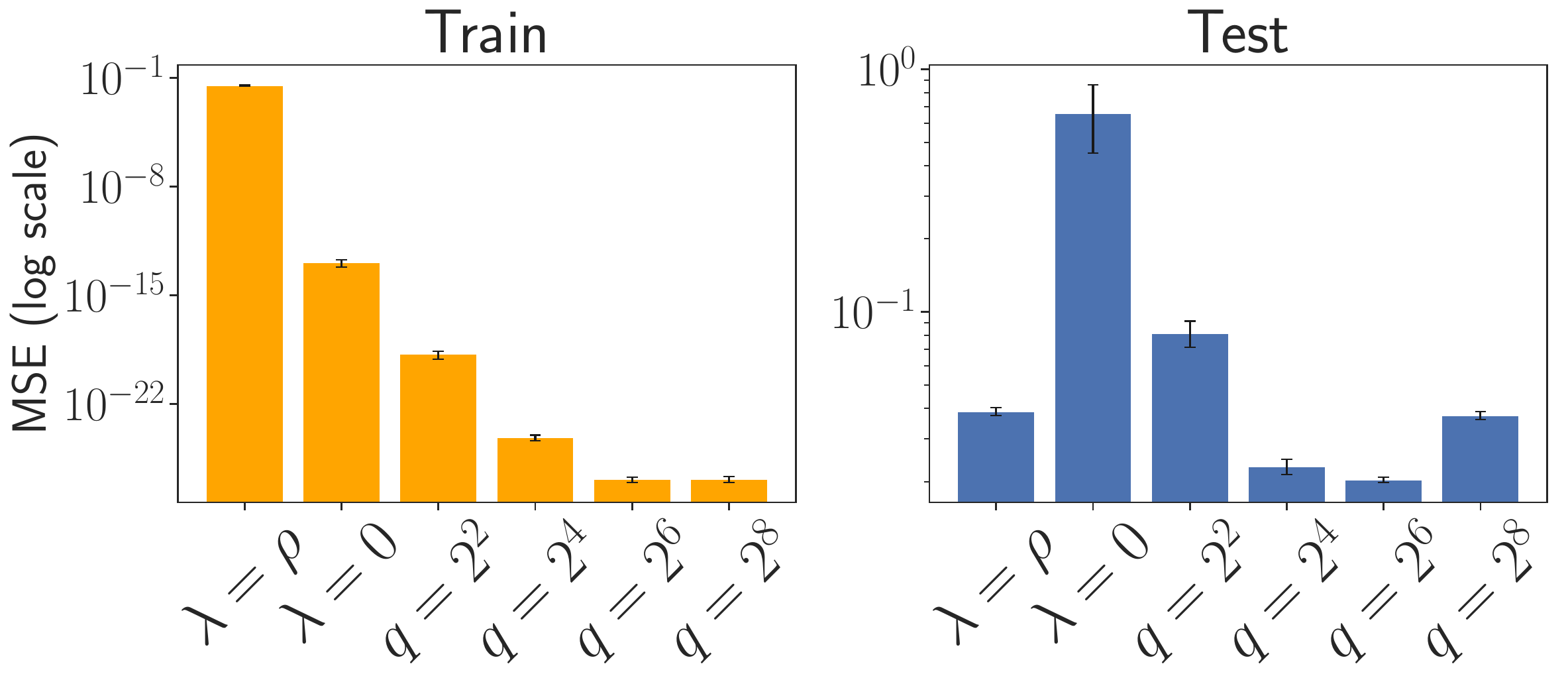}
    \caption{Synthetic regression in $d=7$ (Experiment~(b2 \ref{par:b2}), $n_{\text{train}}=500$). 
Train and test MSE for $k_B$ (ridgeless $\lambda=0$ and ridge $\lambda=\rho$) and ridgeless $k_{LG}$ as $q$ increases. The target is generated from the Local-Global teacher in~\eqref{eq:f3}, so $f^\star \notin \mathcal{H}_{k_B}$.
}
\end{figure}

\medskip
\noindent
\textbf{(b1) Well-specified case with respect to the base kernel.} 
\label{par:b1}
As reported in Figure ~\ref{fig:synthetic_d7in}, the target $f_2$ is generated using the base kernel $k_B$, hence $f_2 \in \mathcal{H}_{k_B}$. In this regime, ridge regression with $k_B$ provides a strong reference solution. As $q$ increases, the interpolating Local-Global estimator approaches the ridge performance while preserving exact interpolation.

\medskip
\noindent
\textbf{(b2) Misspecified case with respect to the base kernel.}
\label{par:b2}
The target $f_3$ is generated using the Local-Global kernel defined in~\eqref{eq:f3}, so that $f_3 \notin \mathcal{H}_{k_B}$. 
As $q$ increases, the test error of the Local-Global estimator approaches that of ridge regression based on $k_B$, in accordance with the theoretical construction. 
In this misspecified regime, however, ridge regression with $k_B$ is not optimal. Intermediate values of $q$ yield strictly lower test error than both ridgeless regression with $k_B$ and the large-$q$ (ridge-like) limit, while remaining in the interpolating regime.

This observation motivates Experiment~(d) on real datasets: if the best ridge estimator based on the base kernel is suboptimal due to misspecification, a Local-Global extension with a suitable value of $q$ may achieve improved generalization, while being in the interpolating regime. Figure~\ref{fig:airfoiln500bestq} illustrates such a scenario.

\paragraph{Experiment (c).}
This experiment examines the asymptotic behavior of the Local-Global estimator as the training size increases. We use the same target $f_2$ as in Experiment~(b1) (see~\ref{par:b1}), defined in~\eqref{eq:target-general}-\eqref{eq:f2}, keeping the anchor points $\{z_j\}$ and coefficients $\{\alpha_j\}$ fixed across all values of $n$. Thus, the underlying function remains unchanged; only the training sample size varies.

For each $n$ listed in Table~\ref{tab:syntheticexperiments}, we generate a new training set of size $n$ together with an independent test set, and compute the mean squared error averaged over $10$ independent repetitions. The parameters are scaled with $n$ as specified in Table~\ref{tab:syntheticexperiments}.

Figure~\ref{fig:consistency} shows that the Local-Global estimator using $q_n$ exhibits decreasing test error that approaches the noise level, illustrating the benign overfitting behavior predicted by the theory. For comparison, ridgeless regression with the base kernel $k_B$ does not display this convergence, as its test error grows with $n$.

\subsection{Real-World data}

We consider two complementary real-world evaluations. Experiment~(d) is a
controlled small-sample study designed to reproduce the behavior observed in
Section~\ref{par:b}: after tuning the local ridge baseline to obtain the best
local-kernel representation of the data, we sweep the exponent $q$ of the
Local-Global kernel and analyze the resulting evolution of the test error.
Experiment~(e) follows a standard validation-based protocol on larger training
sets, with the goal of testing whether the ridgeless Local-Global estimator
improves over the ridgeless local estimator when performance is measured on an
independent test set.

\begin{table}[h]
\centering
\caption{
Summary of the setting of the Experiment~(d). Here \(d\) is the input dimension, \(n_{\mathrm{train}}\)
and \(n_{\mathrm{test}}\) are the training and test sizes, and
\((c,\rho)\) are first tuned for the local ridge baseline.
}
\begin{tabular}{llcccccc}
\toprule
Experiment & Dataset & $d$ & $n_{\mathrm{train}}$ & $n_{\mathrm{test}}$ & $\rho$ & $q$ & $c$ \\
\midrule
d1
& Airfoil
& $20$
& $500$
& $6900$
& tuned
& $\{2^p: p \in \{1,\dots,10\}\}$
& tuned
\\
d2
& cal\_housing
& $8$
& $500$
& $20140$
& tuned
& $\{2^p: p \in \{1,\dots,10\}\}$
& tuned
\\
\bottomrule
\end{tabular}
\label{tab:realexperimentsmall}
\end{table}

\begin{table}[t]
\centering
\caption{Dataset splits used in Experiment~(e).}
\begin{tabular}{lrrrrr}
\toprule
Dataset & $n$ & $d$ & $n_{\mathrm{train}}$ & $n_{\mathrm{val}}$ & $n_{\mathrm{test}}$ \\
\midrule
cal\_housing & 20640 & 8  & 4128 & 6604 & 9908 \\
energy       & 14980 & 14 & 2996 & 4793 & 7191 \\
airfoil      & 7400  & 20 & 1480 & 2368 & 3552 \\
\bottomrule
\end{tabular}
\label{tab:metadata}
\end{table}

\paragraph{Experiment (d).}
This experiment provides a real-data analogue of the synthetic setting in (b), where $q$ interpolates between ridgeless regression and the ridge-like limit induced by the Local-Global construction.
For each dataset, we first tune the bandwidth $c$ and ridge parameter $\rho$ for the local ridge estimator using repeated train/validation splits. Each tuning split uses $n_{\text{train}}$ samples for training and $0.5\cdot n_{\text{train}}$ additional samples for validation. The pair $(c^\star,\rho^\star)$ minimizing the average validation MSE over several such splits is selected.

\begin{figure}[t]
    \centering

    \begin{subfigure}[t]{0.48\linewidth}
        \centering
        \includegraphics[width=\linewidth]{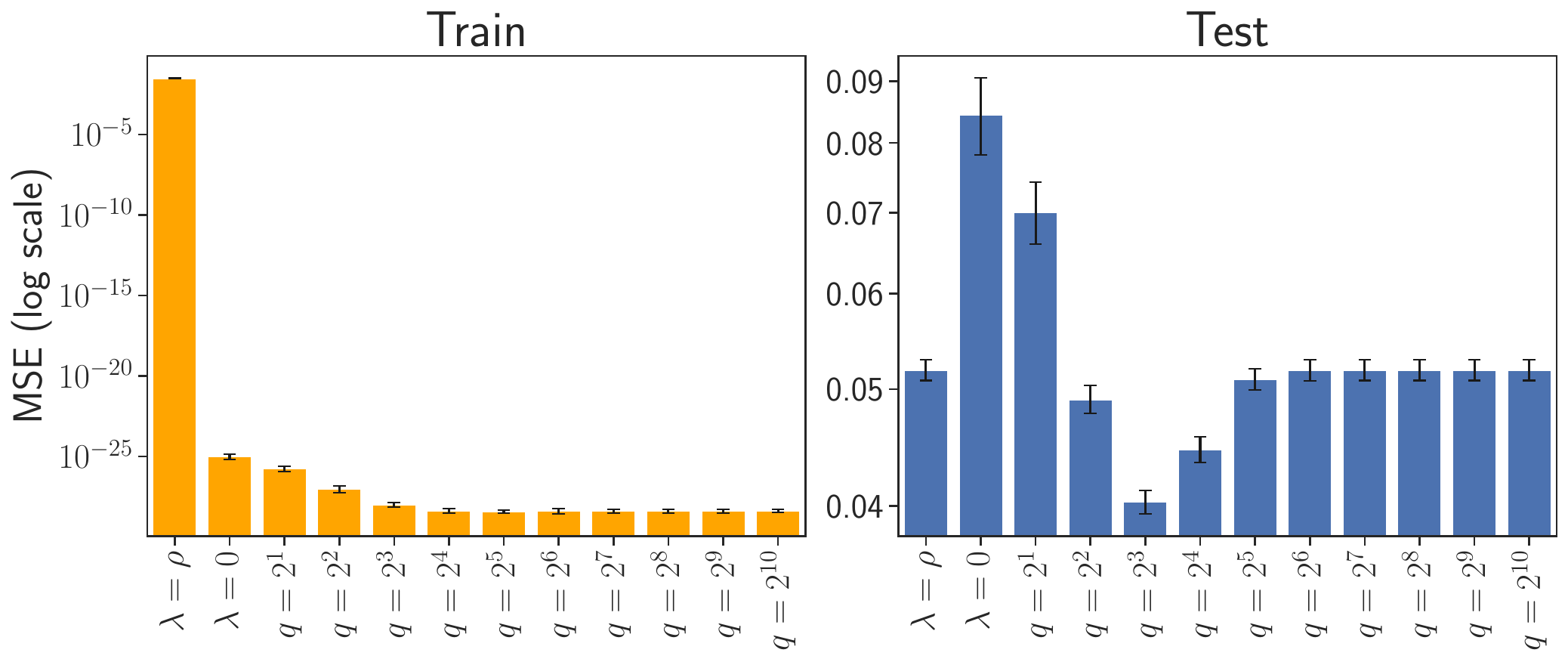}
        \caption{Airfoil - full $q$ sweep}
        \label{fig:airfoil_full}
    \end{subfigure}
    \hfill
    \begin{subfigure}[t]{0.48\linewidth}
        \centering
        \includegraphics[width=\linewidth]{airfoiln500_bestq.pdf}
        \caption{Airfoil - best $q$}
        \label{fig:airfoil_bestqannexe}
    \end{subfigure}

    \vspace{0.5cm}

    \begin{subfigure}[t]{0.48\linewidth}
        \centering
        \includegraphics[width=\linewidth]{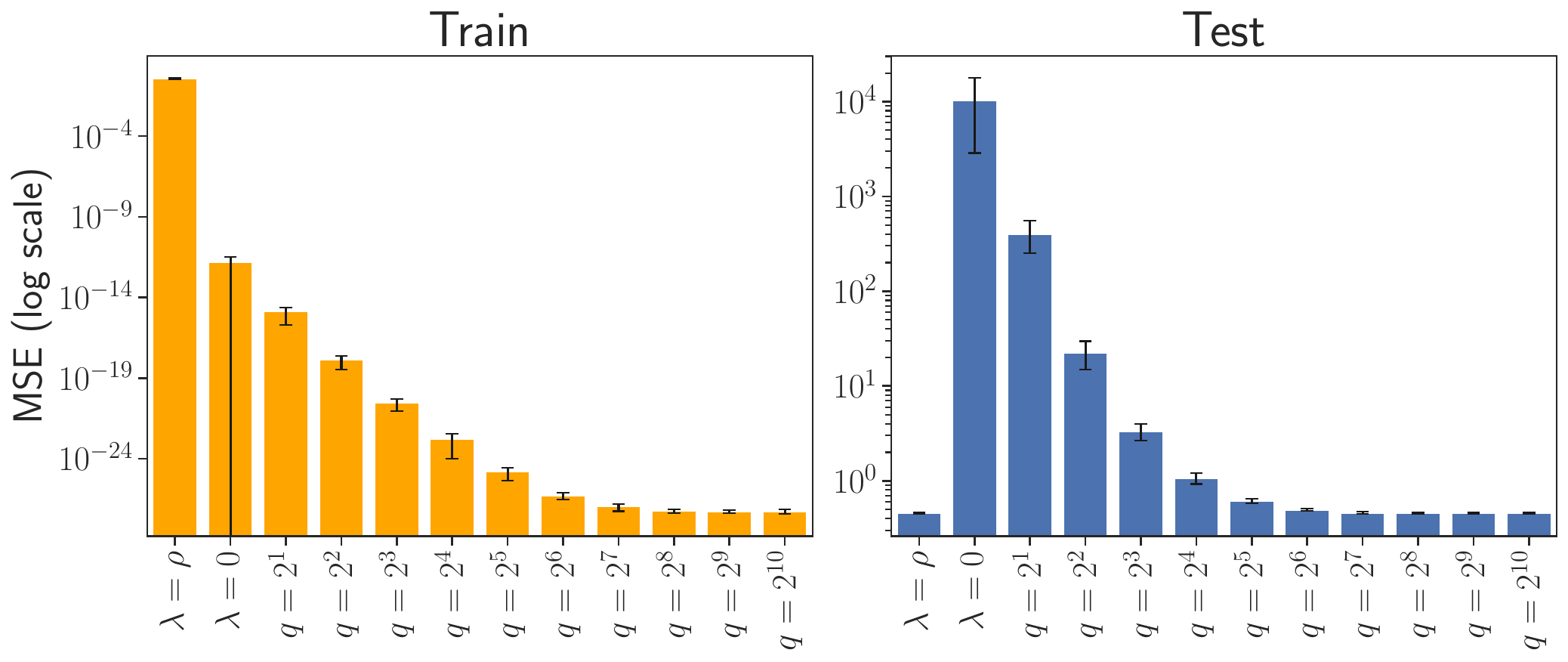}
        \caption{California Housing - full $q$ sweep}
        \label{fig:calhousing_full}
    \end{subfigure}
    \hfill
    \begin{subfigure}[t]{0.48\linewidth}
        \centering
        \includegraphics[width=\linewidth]{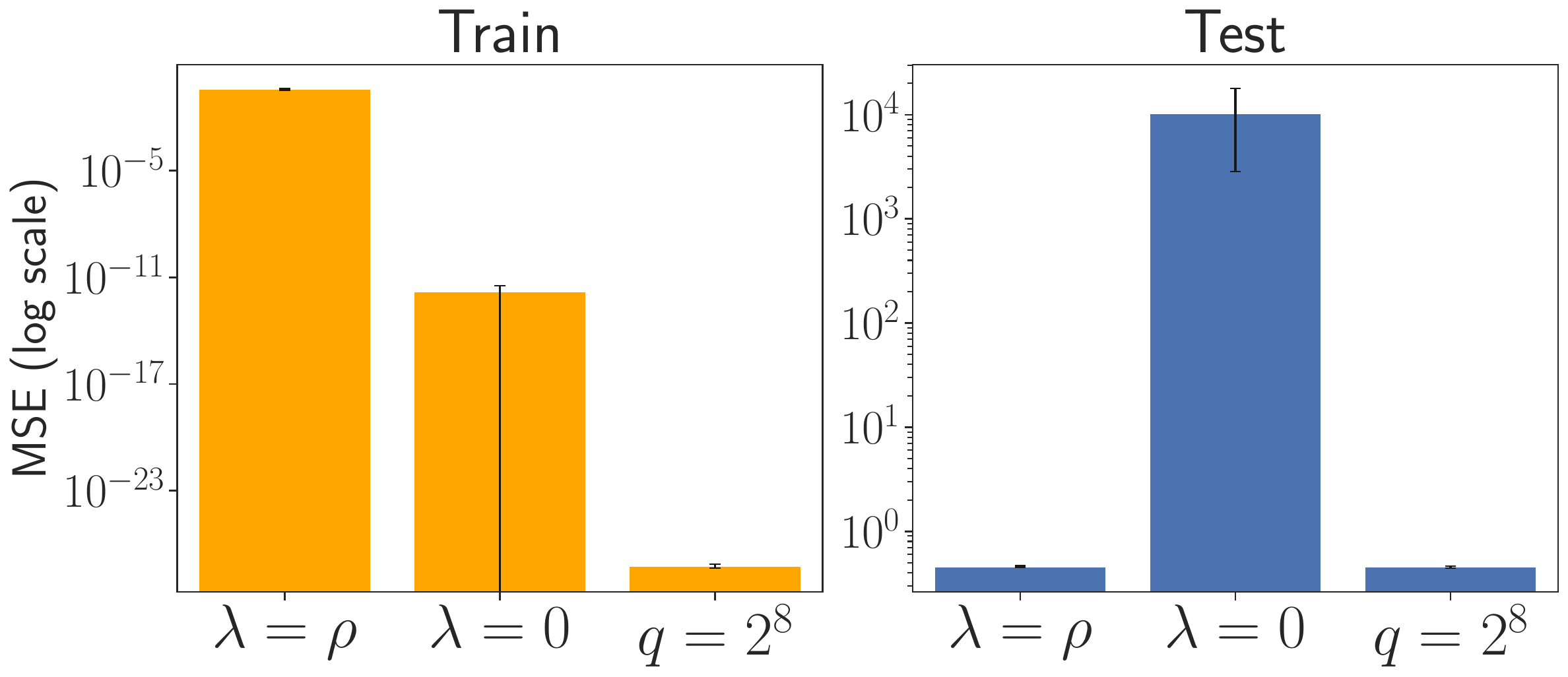}
        \caption{California Housing - best $q$}
        \label{fig:calhousing_bestq}
    \end{subfigure}
    \caption{
    Real-data experiments (Experiment~(d)). 
    Top: Airfoil dataset. Bottom: California Housing dataset. 
    Left: full evolution with $q$. Right: best-performing Local-Global model.
    }
    \label{fig:realdata_grid}
\end{figure}

We then fix $(c^\star,\rho^\star)$ and evaluate on independent train/test splits of size depicted in Table~\ref{tab:realexperimentsmall}:
(i) ridgeless regression with $k_B$,
(ii) ridge regression with $k_B$ using $\lambda=\rho^\star$, and
(iii) ridgeless regression with $k_{LG}$ for the values of $q$ given in Table~\ref{tab:realexperimentsmall}.
Train and test MSE are averaged over $10$ random splits.

The two datasets exhibit behaviors consistent with the two synthetic regimes of Experiment~(b): for Airfoil (d1), intermediate values of $q$ can outperform the tuned local ridge baseline (as in the misspecified pattern of (b2)); for California Housing (d2), the best performance is attained near the ridge-like limit (as in (b1)). These results motivate the practitioner-oriented comparison in the next set of experiments.

\paragraph{Experiment~(e): validation-based real-data evaluation.}
We finally evaluate the proposed Local-Global construction under a more
standard real-data protocol on the regression datasets listed in
Table~\ref{tab:metadata}. All features are standardized to zero mean and
unit variance.

For each dataset, we perform $n_{\text{splits}}=5$ independent random splits. 
In each split, $20\%$ of the samples are used for training. 
The remaining data are partitioned into a validation set (40\% of the remainder) and a test set (60\% of the remainder). 
Model selection is carried out on the validation set, and performance is reported on the independent test set.

\paragraph{Model selection.}
We compare two families of interpolating estimators.

\begin{itemize}
    \item \textbf{Local kernel $k_B$.}  
    The bandwidth $c$ is selected by minimizing the validation MSE over the grid 
    $c_{\mathrm{base}} \in \{0.1,\,0.2,\,0.4,\,0.7,\,1.0\}$. 
    The estimator is trained in the nominally ridgeless regime; however, we add a small numerical regularization $\lambda_{\epsilon} = 10^{-10}$.

    \item \textbf{Local-global kernel $k_{LG}$.}  
    The hyperparameters $(c,\rho,q)$ are jointly selected by minimizing the validation MSE over the grid
    \[
    c \in \{0.1,\,0.2,\,0.4,\,0.7,\,1.0\}, 
    \quad 
    \rho \in \{0.01,\,0.1,\,0.5,\,1.0,\,10\}, 
    \quad 
    q \in  \{3,4,5,6,7,8,9,10\}.
    \]
    After selection, the estimator is refit on the training set using the chosen parameters and evaluated on the test set.
\end{itemize}

For each split, we record both training and test MSE. 
Results are reported as mean $\pm$ standard deviation across splits. 

\begin{table}[h]
    \centering
        \caption{
    Training MSE (mean $\pm$ standard deviation over $5$ splits) for the local kernel $k_B$ and the Local-Global kernel $k_{LG}$ in the ridgeless regime. 
    }
    \input\begin{tabular}{llll}
    \toprule
    Dataset & $k_B$ MSE train & $k_{LG}$ MSE train \\
    \midrule
    airfoil 
    & $(1.3 \cdot 10^{-19}) \pm (6.0 \cdot 10^{-20})$ 
    & $(1.4 \cdot 10^{-28}) \pm (1.0 \cdot 10^{-28})$ \\
    
    cal\_housing 
    & $(2.5 \cdot 10^{-1}) \pm (2.0 \cdot 10^{-2})$ 
    & $(1.1 \cdot 10^{-23}) \pm (6.0 \cdot 10^{-24})$ \\
    
    energy 
    & $(8.7 \cdot 10^{-2}) \pm (3.0 \cdot 10^{-2})$ 
    & $(2.1 \cdot 10^{-22}) \pm (7.0 \cdot 10^{-23})$ \\
    \bottomrule
    \end{tabular}
    \label{tab:trainrr}
\end{table}

\begin{table}[h]
    \centering
        \caption{
Test MSE (mean $\pm$ standard deviation over $5$ splits) for the local kernel $k_B$ and the Local-Global kernel $k_{LG}$. 
    }
    
    \begin{tabular}{llll}
    \toprule
    Dataset & $k_B$ MSE test & $k_{LG}$ MSE test \\
    \midrule
    airfoil 
    & $(6.0 \cdot 10^{-2}) \pm (6.02 \cdot 10^{-3})$ 
    & $(3.0 \cdot 10^{-2}) \pm (1.45 \cdot 10^{-3})$ \\
    
    cal\_housing 
    & $(2.56 \cdot 10^{3}) \pm (3.76 \cdot 10^{3})$ 
    & $(8.0 \cdot 10^{-1}) \pm (3.0 \cdot 10^{-2})$ \\
    
    energy 
    & $(5.26 \cdot 10^{2}) \pm (7.7 \cdot 10^{3})$ 
    & $(1.18 \cdot 10^{0}) \pm (8.0 \cdot 10^{-2})$ \\
    \bottomrule
    \end{tabular}
    \label{tab:testrr}
\end{table}

Tables~\ref{tab:trainrr} and~\ref{tab:testrr} report the predictive performance of the two kernel families. The Local-Global estimator systematically operates in the interpolating regime, achieving near-zero training error across splits. In contrast, the local baseline does not always interpolate and exhibits noticeable training error on datasets such as California Housing and Energy.

On the test set, both methods operate in regimes that may be associated with overfitting. 
However, the nature of this overfitting differs. 
The local baseline exhibits poor and unstable generalization despite achieving low training error. 
By contrast, the interpolating Local-Global estimator maintains stable performance across splits (low standard deviation) together with improved test accuracy. 

Table~\ref{tab:bestparamrr} lists the hyperparameters achieving the lowest test error among the evaluated configurations.

\begin{table}[h]
    \centering
     \caption{
    Hyperparameters achieving the lowest test MSE across splits for each dataset. 
    For $k_B$, we report the best bandwidth $c_{\mathrm{base}}^*$. 
    For $k_{LG}$, we report the best $(c^*, \rho^*, q^*)$ together with the corresponding test error.
    }
    \begin{tabular}{lrrrrr}
    \toprule
    Dataset & $n$ & $d$ & $n_{\mathrm{train}}$ & $n_{\mathrm{val}}$ & $n_{\mathrm{test}}$ \\
    \midrule
    cal\_housing & 20640 & 8 & 4128 & 6604 & 9908 \\
    energy & 14980 & 14 & 2996 & 4793 & 7191 \\
    airfoil & 7400 & 20 & 1480 & 2368 & 3552 \\
    \bottomrule
    \end{tabular}
    \label{tab:bestparamrr}
\end{table}

\section{Local-Global Quantum Kernels as Quantum Embedding Kernel}
\label{app:qemb}

In this appendix, we show that the Local-Global quantum kernel can be interpreted as a standard quantum embedding kernel associated with a single unitary feature map acting on an enlarged Hilbert space.

Let $U(x)$ be a unitary quantum feature map acting on $t$ qubits. To obtain the normalization we consider the case where $\sqrt{\lambda_L} + \sqrt{\lambda_G} =1$ with $\lambda_L,\lambda_G\ge0$.   

Let $\rho_{\mathrm{init}}^{L}$ and $\rho_{\mathrm{init}}^{G}$ denote the initial states associated with the local and global kernels, see Definition \ref{de:lgk}, and define the corresponding embeddings 
\begin{equation}
\rho_x^L = U(x)\rho_{\mathrm{init}}^{L}U^\dagger(x),
\qquad
\rho_x^G = U(x)\rho_{\mathrm{init}}^{G}U^\dagger(x).
\end{equation}

We construct a single embedding acting on $t+1$ qubits. Define the enlarged initial state
\begin{equation}
\rho_{\mathrm{init}}^{LG}
=
\sqrt{\lambda_L} \, |0\rangle\langle 0| \otimes \rho_{\mathrm{init}}^{L}
+
\sqrt{\lambda_G}  \, |1\rangle\langle 1| \otimes \rho_{\mathrm{init}}^{G},
\end{equation}
and the unitary
\begin{equation}
W(x) = I_2 \otimes U(x),
\end{equation}
where the first qubit acts as a control register.

The associated quantum embedding is
\begin{equation}
\rho_x^{LG}
=
W(x)\rho_{\mathrm{init}}^{LG}W^\dagger(x).
\end{equation}

By construction,
\begin{equation}
\rho_x^{LG}
=
\sqrt{\lambda_L}  |0\rangle\langle 0| \otimes \rho_x^L
+
\sqrt{\lambda_G}  |1\rangle\langle 1| \otimes \rho_x^G.
\end{equation}

Consider the Hilbert-Schmidt inner product $\mathrm{Tr}\!\left[\rho_x^{LG} \rho_z^{LG}\right].$ Using the orthogonality relation $\mathrm{Tr}\!\left[|0\rangle\langle 0|\,|1\rangle\langle 1|\right] = 0,$ the cross terms vanish and we obtain
\begin{align}
\mathrm{Tr}\!\left[\rho_x^{LG} \rho_z^{LG}\right]
&=
\lambda_L
\mathrm{Tr}\!\left[
|0\rangle\langle 0|
\otimes
\rho_x^L \rho_z^L
\right]
+
\lambda_G
\mathrm{Tr}\!\left[
|1\rangle\langle 1|
\otimes
\rho_x^G \rho_z^G
\right] \\
&=
\lambda_L
\mathrm{Tr}[|0\rangle\langle 0|]
\,\mathrm{Tr}[\rho_x^L\rho_z^L]
+
\lambda_G
\mathrm{Tr}[|1\rangle\langle 1|]
\,\mathrm{Tr}[\rho_x^G\rho_z^G] \\
&=
\lambda_L k_L(x,z)
+
\lambda_G k_G(x,z).
\end{align}

Therefore, the Local-Global kernel can be interpreted as a standard quantum embedding kernel obtained from a single unitary feature map acting on an enlarged Hilbert space. The additional qubit acts as a classical selector between local and global initializations, and the resulting Hilbert-Schmidt inner product exactly reproduces the weighted kernel.

\end{document}